\documentclass[12pt]{article}

\usepackage{epsf}
\usepackage[dvips]{graphicx}

\begin{document}

\begin{center}
{\bfseries $z$-SCALING  AT  RHIC  AND TEVATRON}

\vskip 5mm
M.V.~Tokarev$^{1, \natural}$, I.~Zborovsk\'{y}$^{2}$\\
T.G.~Dedovich$^{1}$

\vskip 5mm

{\small
(1) {\it Joint Institute for Nuclear Research,
Dubna, Russia }
\\
(2) {\it Nuclear Physics Institute ASCR, \v{R}e\v{z}, Czech Republic }
\\
$^{ \natural}${\it E-mail: tokarev@sunhe.jinr.ru}}
\end{center}

\vskip 5mm

\begin{center}
\begin{minipage}{150mm}
\centerline{\bf Abstract}
The experimental data on inclusive cross sections of jet,
direct photon and hadron production in $pp/\bar{p}p$ and $AA$ collisions
at RHIC and Tevatron are analyzed in the framework of $z$-scaling.
Results of analysis are compared with data obtained
at ISR, S$\rm \bar p$pS and Tevatron.
The properties of $z$-presentation of experimental data are verified.
Physical interpretation of the scaling function $\psi(z)$ and variable $z$
is discussed. The locality, self-similarity and
fractality are argued to reflect the general structure of the
colliding objects, interaction of their constituents and particle
formation at small scales.
The obtained results suggest that the $z$-scaling may be used as a tool
for searching for new physics phenomena beyond Standard Model
in hadron and nucleus collisions at high transverse momentum
and high multiplicity at U70, RHIC, Tevatron and LHC.
\end{minipage}
\end{center}


{\section{Inroduction}}

Study of particle production at large transverse momenta in
high energy collisions of hadrons and nuclei
is of interest to search for exotic phenomena such as quark compositeness
\cite{Quarkcom}, extra dimensions \cite{Extradim}, black holes
\cite{Blackhole},  fractal space-time \cite{Fracspace}  and
collective phenomena such as phase transition
of nuclear matter and formation of quark-gluon plasma \cite{RHIC_review}.
New features of particle formation experimentally established
could be crucial for precise test of Quantum chromodynamics (QCD)
and Electroweak (EW) theories in perturbative and non-perturbative regimes.
Many phenomenological approaches were suggested  for description
of regularities reflecting general properties
(locality, self-similarity) of hadron and nucleus interactions
at a constituent level at high energies \cite{Feynman}-\cite{Brodsky}.

In the present paper we use the concept of the $z$-scaling
\cite{Z}-\cite{ZZ}
for analysis of new experimental data on inclusive spectra of hadron,
direct photon and jet production
in $pp/\bar{p}p$ and $AA$ collisions at RHIC and Tevatron.
We show that the $z$-scaling represents regularity in both
soft and hard regime of particle production over a wide kinematical
range for events with different centrality.
The procedure for construction of
the scaling function $\psi(z)$ and scaling variable $z$
for different types of produced objects is described.
Both quantities are expressed via the experimentally measured
inclusive cross section $Ed^3\sigma/dp^3$, the multiplicity
density $dN_{\rm ch}/d\eta$ and kinematical characteristics
of colliding and produced particles.
Properties of $z$-scaling are used to predict
particle spectra of $J/\psi, D^0, B^+ $ mesons and $Z, W^+$ bosons
at higher collision energies and transverse momenta.
We suggest to use the
$z$-scaling as a tool for searching for new physics phenomena
of particle production in high transverse momentum and high
multiplicity region at U70, Tevatron, RHIC and LHC.

{\section{$z$-Scaling}}

Here we would like to remind basic ideas and some formulas
which are important for understanding
of the developed approach \cite{Z}-\cite{ZZ}.

The idea of $z$-scaling is based on the assumption \cite{Stavinsky}
that gross features of inclusive particle distribution
of the reaction written  in the symbolic form as follows
\begin{equation}
M_{1}+M_{2} \rightarrow m_1 + X
\label{eq:r1}
\end{equation}
can be described at high energies in terms of the corresponding
kinematic characteristics of constituent binary collision
\begin{equation}
(x_{1}M_{1}) + (x_{2}M_{2}) \rightarrow m_1/y +
(x_{1}M_{1}+x_{2}M_{2} + m_2/y)
\label{eq:r2}
\end{equation}
Here $x_1$ and $x_2$ are the fractions of the incoming 4-momenta $P_1$ and $P_2$
of the objects with the masses $M_1$ and $M_2$ carried out by constituents.
The inclusive particle with the mass $m_1$ and the 4-momentum $p$
carries out the fraction $y$ of the 4-momentum of the outgoing constituent.
The parameter $m_2$ is introduced to satisfy the internal conservation laws
(for baryon number, isospin, strangeness, and so on). It is determined from the
corresponding exclusive reaction.
For example, the parameter $m_2$
 for the exclusive process
 $p+p\rightarrow \pi^++(p+n)$ is equal to $m_n-n_p$ as follows from the relation
  $\pi^++(p+n)=\pi^++p+p+(n-p)$ satisfying to the baryon
   and electric charge conservation laws.
The parameters for inclusive $K^+,\Lambda^0$ and $\bar p$ \
production in $p+p$ collisions are equal to $m_{\Lambda}-m_p$,
$m_K-m_p$ and $m_p$, respectively.
It is assumed that the constituent interaction satisfies
the energy-momentum conservation law written in the form
\begin{equation}
(x_1P_1+x_2P_2-p/y)^2 =(x_1M_1+x_2M_2+m_2/y)^2.
\label{eq:r3}
\end{equation}
The equation expresses locality of hadron interaction
at constituent level.

The scaling variable $z$ is defined as follows \cite{ZZ}~{\footnote{Other modifications
of $z$ are described in \cite{Z,Z3,ZS,ZZZ}}.}
\begin{equation}
z = \frac{s^{1/2}_{\bot}}{(dN/d\eta|_0)^c \cdot m_0}\cdot\Omega^{-1}.
\label{eq:r17}
\end{equation}
Here $m_0$ is a mass constant which is fixed at the value of nucleon
mass, $dN/d\eta|_0$ is a multiplicity
density at (pseudo)rapidity $\eta=0$,
the parameter $c$ has physical meaning of "specific heat" of produced medium.
 The quantity $s^{1/2}_{\bot}$ is a minimal transverse kinetic energy
 of the constituent sub-process.
 It consists of two parts $s^{1/2}_{\lambda}$ and $s^{1/2}_{\chi}$
which represent the energy for creation of the inclusive particle
and its recoil, respectively.
The quantity $\Omega$ depends on the momentum fractions $x_1,x_2,y$
and the anomalous fractal dimensions $\delta_1$, $\delta_2$ and $\epsilon$
as follows
\begin{equation}
\Omega(x_1,x_2,y)=(1-x_1)^{\delta_1}(1-x_2)^{\delta_2}(1-y)^{\epsilon}.
\label{eq:r6}
\end{equation}
It is proportional to relative number of parton configurations containing
constituents which carry the fractions $x_1$ and  $x_2$ of the incoming momenta
$P_1$ and $P_2$  and the outgoing constituent which fraction $y$ is carried out
by the inclusive particle with the momentum~$p$.

The fractions are determined in a way to minimize  $\Omega^{-1}(x_{1},x_{2},y)$
taking into account the energy-momentum conservation law
in the binary collision (\ref{eq:r2}).
This is equivalent to the solution of the system of nonlinear equations
\begin{equation}
{\partial\Omega(x_1,x_2,y)} / {\partial x_1} = 0, \ \ \ \ \ \
{\partial\Omega(x_1,x_2,y)} / {\partial x_2} = 0, \ \ \ \ \ \
{\partial\Omega(x_1,x_2,y)} / {\partial y} = 0
\label{eq:r8}
\end{equation}
with the additional condition (\ref{eq:r3}).

The scaling function $\psi(z)$ is expressed in terms of the experimentally
measured inclusive invariant cross section $Ed^3\sigma/dp^3$, multiplicity
density $dN/d\eta$, total inelastic cross section $\sigma_{in}$
and kinematical variables (masses and momenta)
 characterizing the inclusive process.
It can be written as follows
\begin{equation}
\psi(z) = -{ { \pi s} \over { (dN/d\eta) \sigma_{in}} } J^{-1} E {
{d^3\sigma} \over {dp^3}  }.
\label{eq:r21}
\end{equation}
Here, $s$ is the center-of-mass collision energy squared and
$J$ is the corresponding Jacobian.
The function $\psi(z)$ determined by (\ref{eq:r21}) satisfies
to the normalization condition
\begin{equation}
\int_{0}^{\infty} \psi(z) dz = 1.
\label{eq:r24}
\end{equation}
The relation allows us to interpret the $\psi(z)$ as a
probability density to produce inclusive particle with the corresponding
value of the variable $z$.

{\section{$z$-Scaling in $pp $ and $AA$ collisions at RHIC}}

In this section we analyze experimental data
on particle transverse spectra
obtained in $pp$ and $AA$ collisions at RHIC.

{\subsection{$\pi^0$ mesons}}

The PHENIX  and STAR collaborations published the new data
\cite{pi0_PHENIX,pi0_STAR} on the inclusive spectrum
of $\pi^0$-mesons produced in $pp$ collisions
in the central rapidity range at energy $\sqrt s = 200$~GeV.
The transverse momenta of $\pi^0$-mesons
 were measured up to 20~GeV/c.
 The $p_T$ and $z$ presentations of data for $\pi^0$-meson spectra
 obtained at ISR  \cite{Angel}-\cite{Eggert}
 and RHIC \cite{pi0_PHENIX,pi0_STAR} are shown in Figs. 1(a) and 1(b).
One can see that
the PHENIX and STAR data are compatible each other over a overlapping range.
The  $p_T$-spectra of $\pi^0$-meson production
demonstrate the strong dependence on collision energy.
As seen from Fig. 1(b) the new data on $\pi^0$-meson inclusive
cross sections obtained at the RHIC
are in a good agreement  with our earlier results \cite{Zpi0}.
The shape of the scaling functions for ISR and RHIC energies
is observed to be the same.
Uncertainty of relative normalization factor for cross sections
is found to be 2.
Based on the obtained results we conclude that
the available experimental data on high-$p_T$
$\pi^0$-meson production in  $pp$ collisions confirm
the property of the energy independence
of $\psi(z)$ in $z$ presentation.

\begin{figure}
\hspace*{0mm}
\includegraphics[width=70mm,height=70mm]{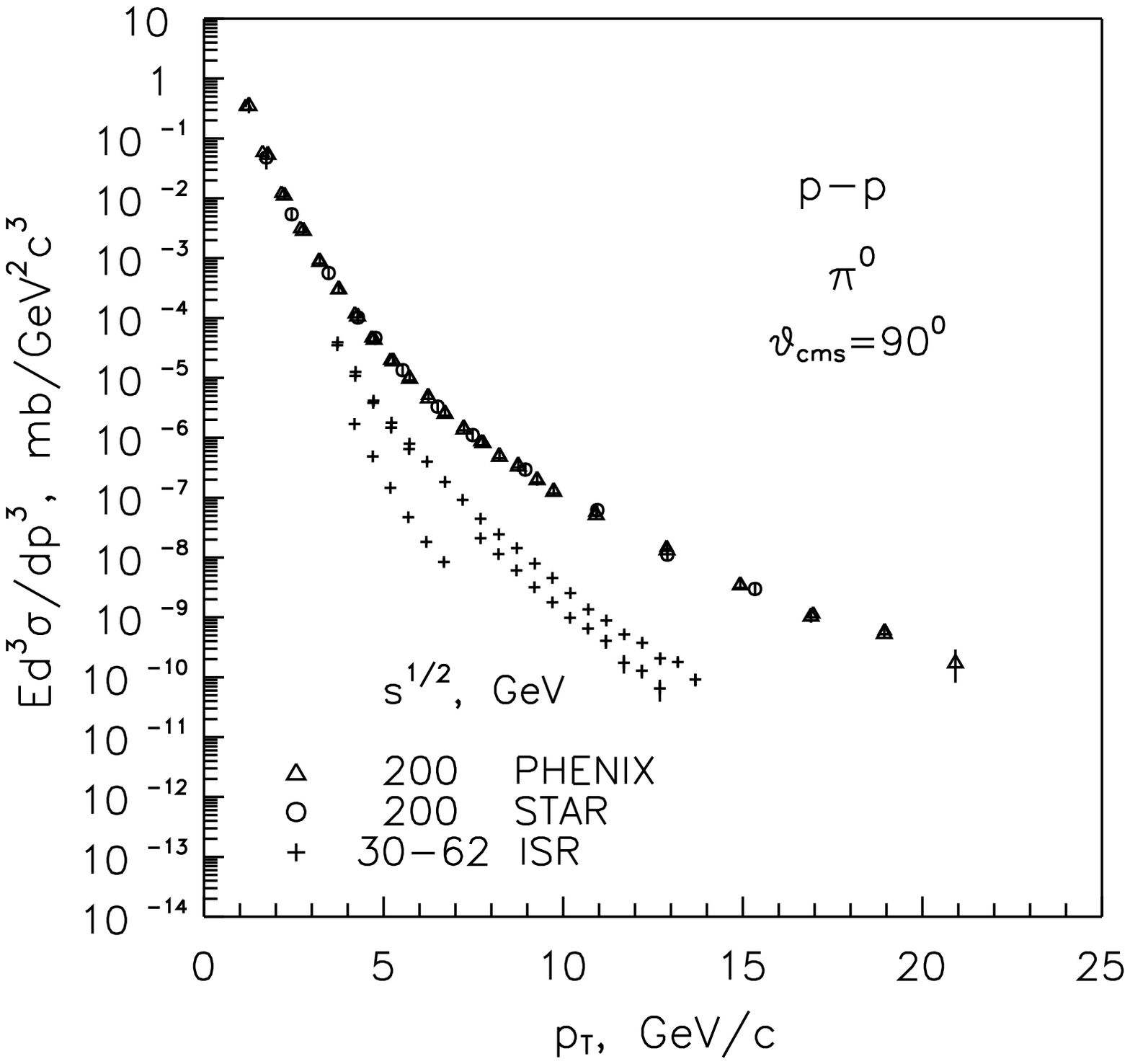}
\hspace*{10mm}
\includegraphics[width=70mm,height=70mm]{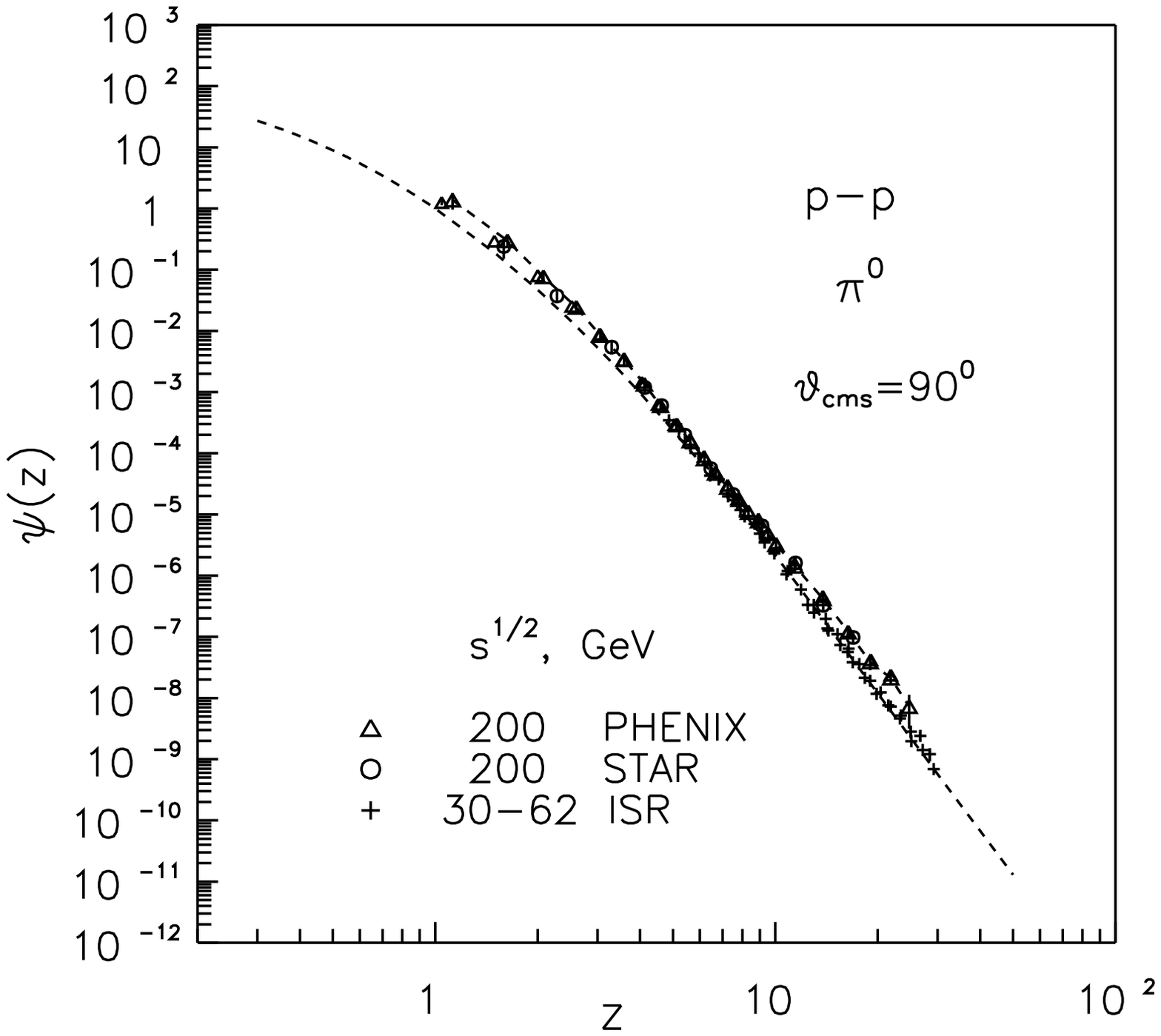}
\hspace*{40mm} (a) \hspace*{80mm} (b)
\caption{ Spectra of $\pi^0$ mesons produced in $pp$ collisions
in $p_T$ and $z$ presentations.
Experimental data are taken from \cite{pi0_PHENIX,pi0_STAR}.}
\end{figure}

{\subsection{Charged hadrons}}

Here we present the results of the joint analysis of experimental data
on charged hadrons produced in $\bar{p}p$  \cite{mult_E735} and $pp$
\cite{mult_STAR} collisions at different multiplicities of charged
particles and  different incident energies over a wide $p_T$
range.

The E735 collaboration measured the multiplicity dependence of
charged hadron spectra \cite{mult_E735}
in  proton--antiproton collisions at the
energy $\sqrt{s}=1800~$GeV for $dN_{\rm ch}/d\eta=2.3-26.2$
at  Tevatron.  These measurements  include highest
multiplicity density per nucleon--(anti)nucleon collision obtained
so far. The pseudorapidity range was $|\eta|<3.25$. The data cover the
transverse momentum range $p_T = 0.15-3~$GeV/c.
The data demonstrate strong
sensitivity of the spectra to the multiplicity density at high $p_T$.
The independence of $\psi$ on multiplicity density $dN_{\rm ch}/d\eta$
was observed. The result gives strong restriction on the parameter
$c$ which was found to be $c=0.25$.

The STAR collaboration obtained the new data \cite{mult_STAR}
on the inclusive spectrum
of charged hadrons produced in proton-proton collisions in the
central rapidity range $|\eta|<0.5$ at the energy $\sqrt s = 200$~GeV at RHIC.
 The transverse momentum spectra were measured up to
9.5~GeV/c. The data demonstrate the strong dependence of the spectra
on the multiplicity
density at $dN_{\rm ch}/d\eta=2.5, 6.0$ and 8.0.
The STAR data confirm the multiplicity independence of the scaling function $\psi(z)$
established for the proton-antiproton collisions at higher
energies.
The corresponding value of the parameter $c$ was found to be the same 0.25.

We would like to note that the scaling for charged particles
produced in proton--antiproton  and proton--proton collisions for
different multiplicities and energies is consistent with the
values of the anomalous fractal dimensions $\delta_1=0.7$,
$\delta_2=0.7$ and $\epsilon=0.7$.
Experimental data on
multiplicity dependence of the spectra for non-identified charged
hadrons obtained by the UA1 \cite{mult_UA1} and CDF \cite{mult_CDF}
collaborations at the S${\rm \bar p}$pS and Tevatron are found to be
in good agrement with the scaling function
at $c=0.25$ similarly as the E735 and STAR data.

The results of $z$-presentation for $\bar{p}p$ and $pp$ collisions
are used to construct the scaling function over a wide range of $z$.
There is indication that in the low $z$-range the shape of the scaling function
for both collisions is the same.
In the same time the asymptotic behavior of $\psi(z)$ for $\bar{p}p$
and $pp$ for high $z$ is different. The difference increases with $z$.
The E735 data in $z$-presentation cover the range $z=3\cdot 10^{-2}-5$.
For high $z$ the STAR data demonstrate power law. In the overlapping range $z=0.4-5$
the behavior of $\psi(z)$ for $\bar{p}p$ and $pp$ collisions coincides each other.
Figure 2 demonstrates the result of combine analysis
of the E735 and STAR data on cross section in $z$ presentation.

{\subsection{Multiplicity dependence of particle spectra}}

We use the properties of $z$-scaling to predict cross sections
of charged hadron production in $pp$ collisions
over a wide range of transverse momenta,
multiplicity densities and collision energies.

Figure 3 shows multiplicity dependence of transverse
spectra of charged particles produced in $pp$ collisions
in central rapidity range at $\sqrt s = 200$ and 11.5~GeV.
The predictions are of interest for searching for phase transition
of hadron matter at extremely high multiplicity (energy) density
and can be verified at U70 and RHIC.
We suppose that violation of $z$-scaling in particle production
is a signature of such phase transition.

\begin{figure}
\hspace*{35mm}
\includegraphics[width=80mm,height=80mm]{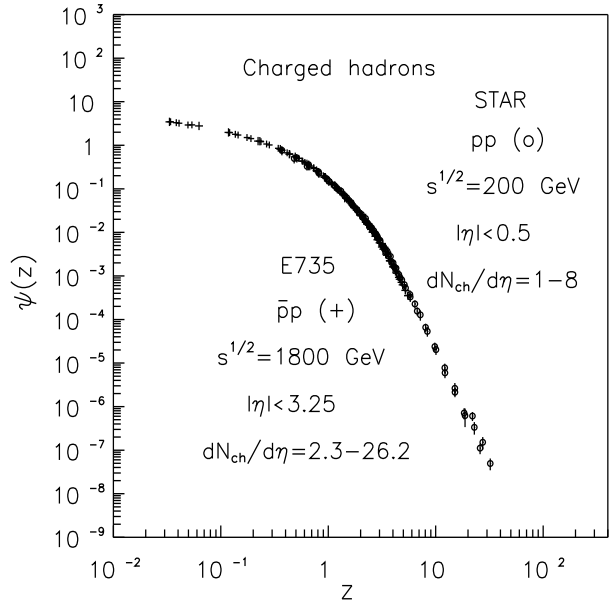}
\caption{Spectra of charged hadrons produced
in $\bar{p}p$ and $pp$ collisions
in $z$ presentations.
Experimental data are taken from \cite{mult_E735,mult_STAR}.}
\end{figure}

\begin{figure}
\hspace*{0mm}
\includegraphics[width=70mm,height=70mm]{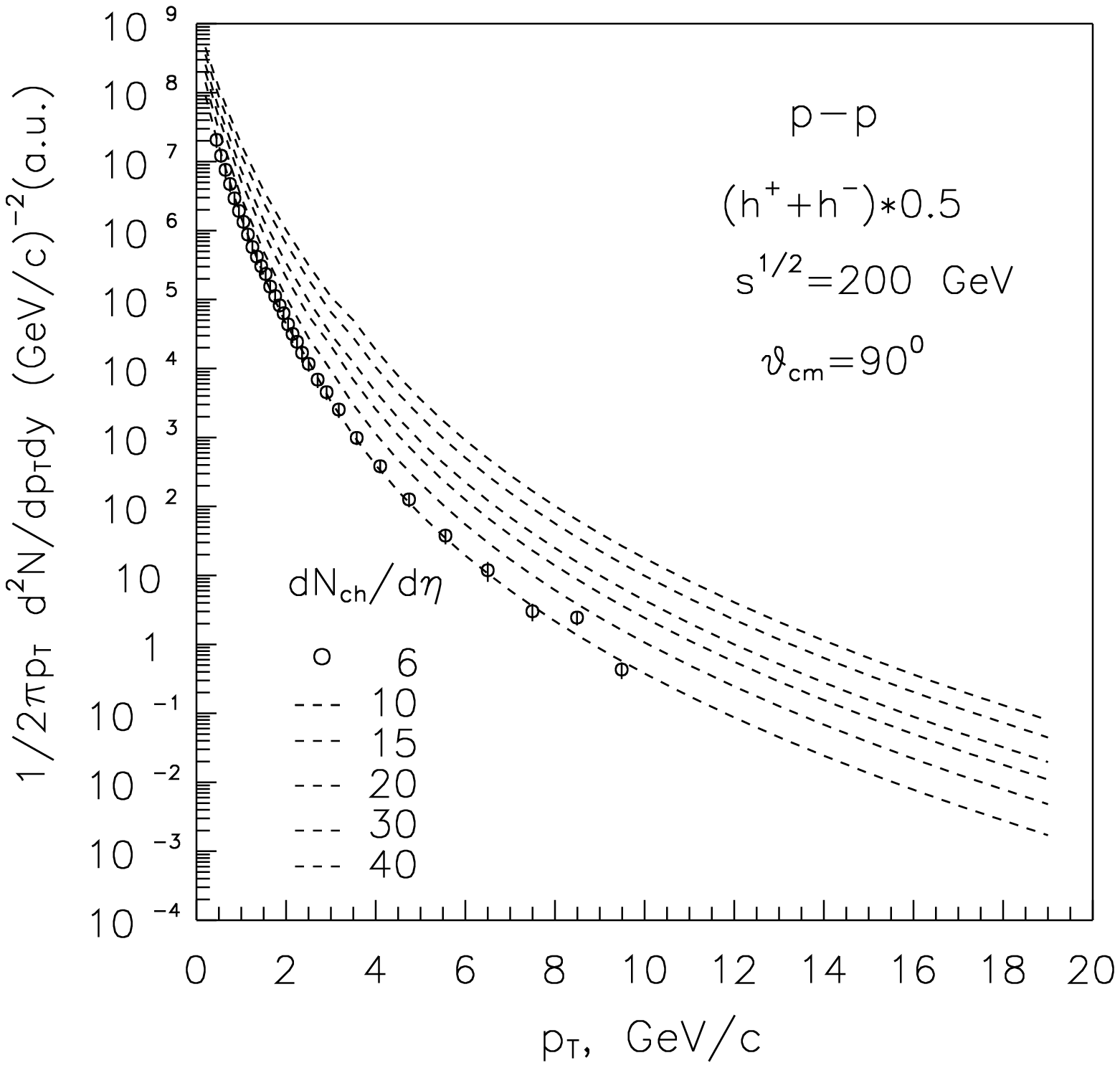}
\hspace*{10mm}
\includegraphics[width=70mm,height=70mm]{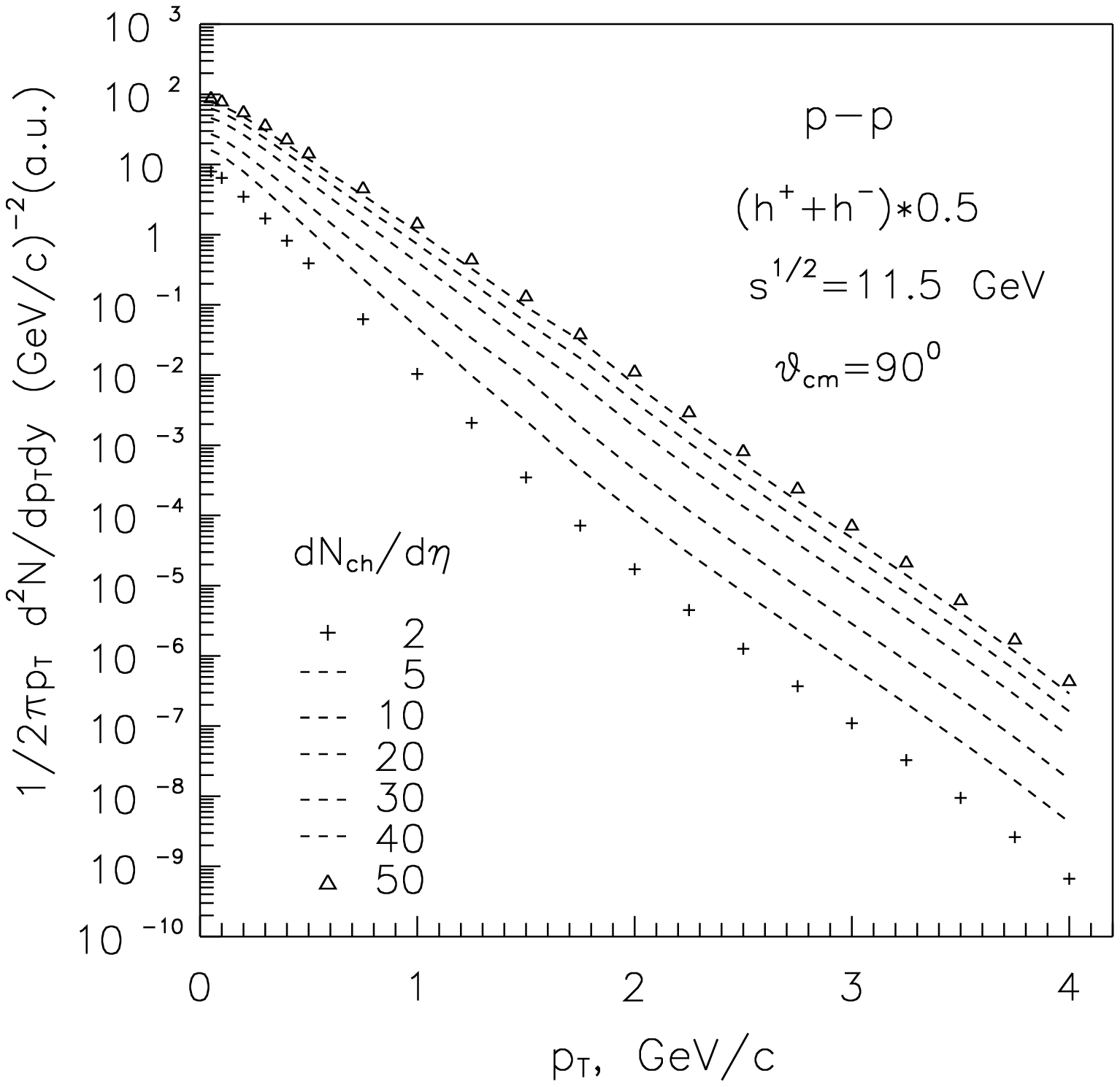}
\hspace*{40mm} (a) \hspace*{80mm} (b)
\caption{Spectra of charged hadrons produced in $pp$ collisions
for different multiplicity density $dN_{ch}/d\eta$
in $p_T$ presentations at RHIC (200~GeV) (a) and U70 (11.5~GeV) (b) energies.
Experimental data are taken from \cite{mult_STAR}.}
\end{figure}

{\subsection{Dependence of $<p_T>$ on multiplicity and collision energy}}

The correlation between $<p_T>$ and multiplicity density
of charged hadrons $dN_{\rm ch}/d\eta$ in high-energy
hadronic collisions was experimentally observed at ISR \cite{SFM},
S$\rm \bar p$pS \cite{mult_UA1} and Tevatron\cite{E735_pT,mult_E735,mult_CDF}.
The dependence of $<p_T>$ on collision energy $\sqrt s$
was experimentally established as well.
It was found that the average $p_T$ grows with multiplicity,
collision energy and mass of produced particle.

\begin{figure}
\hspace*{0mm}
\includegraphics[width=70mm,height=70mm]{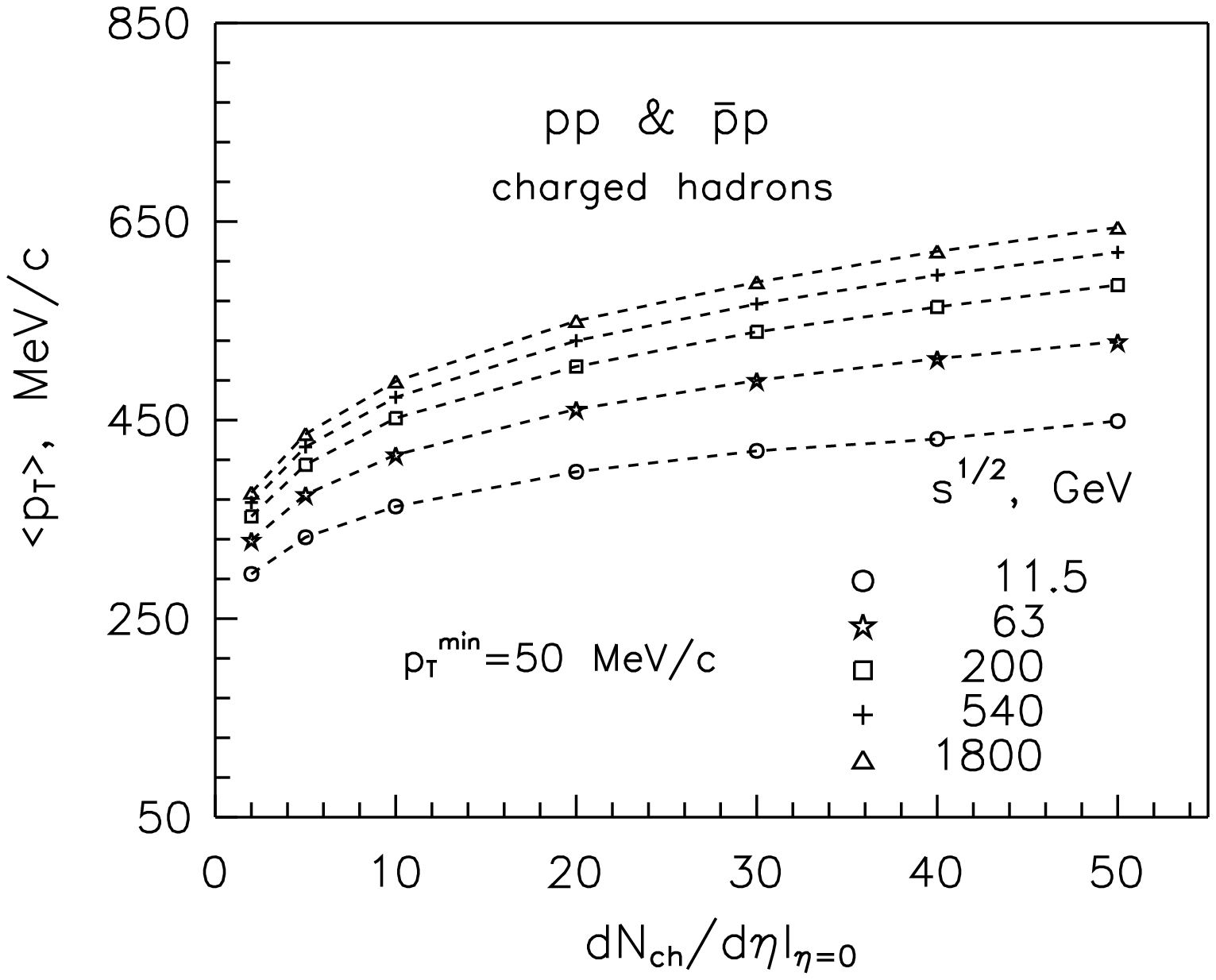}
\hspace*{10mm}
\includegraphics[width=70mm,height=70mm]{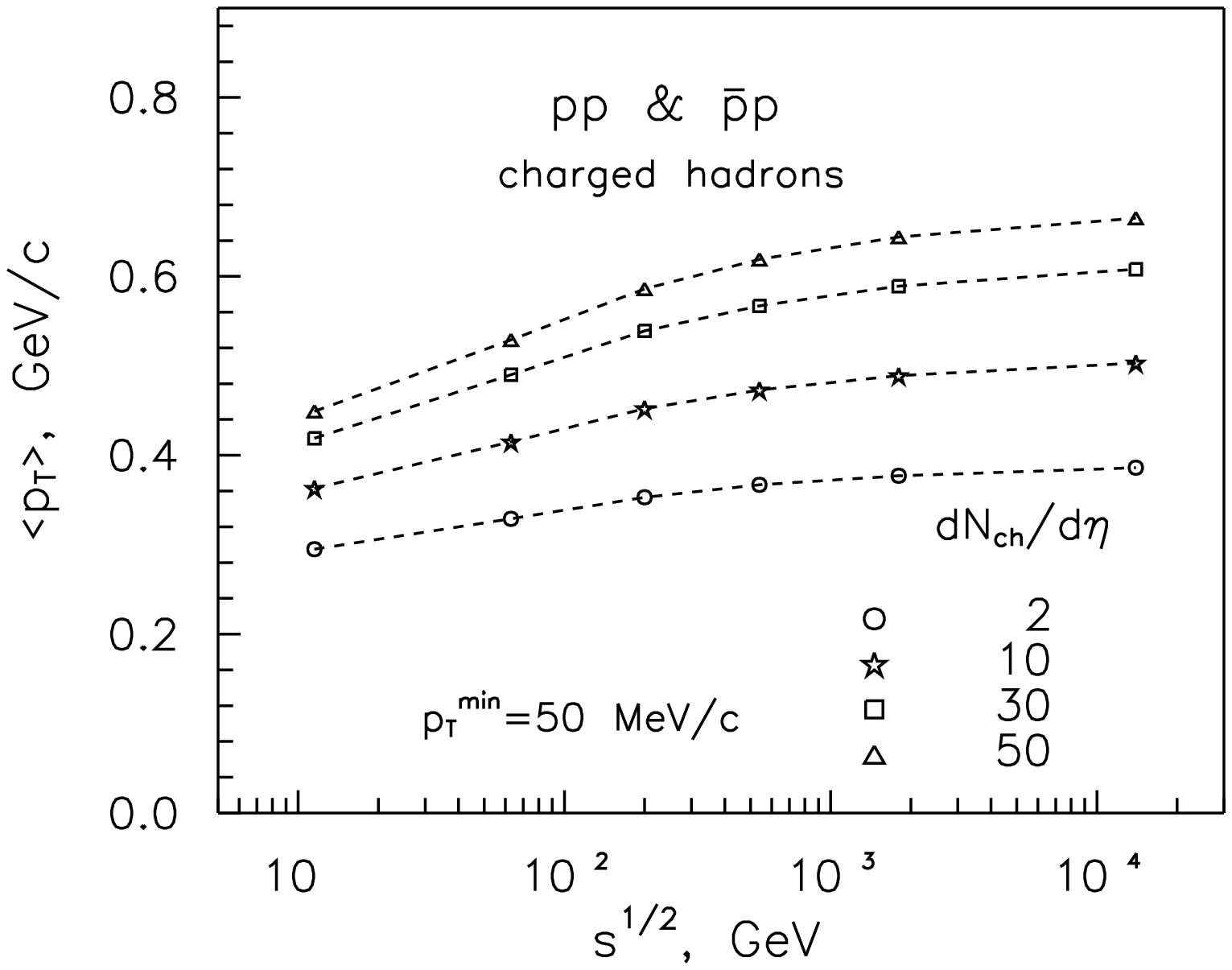}
\hspace*{40mm} (a) \hspace*{80mm} (b)
\caption{Mean transverse momentum $ <p_T> $ as a function
of  multiplicity density $dN_{ch}/d\eta$ (a)
and collision energy $\sqrt s $ (b).}
\end{figure}

It is considered that the mean value of transverse momentum $<p_T>$
characterizes medium created in collisions of hadrons or nuclei.
The  dependence of $<p_T>$ on collision energy
$\sqrt s$ and multiplicity density $dN_{\rm ch}/d\eta$
is a useful tool to investigate the collective behavior
of soft multi-particle production.
It could give indications on phase transition of hadron matter
at extremely high hadron density.

As noted in \cite{mult_CDF} the different theoretical models
used for explanation
of the phenomena (mechanism of multi-particle production)
do not provide satisfactory predictions for existing experimental results
leaving the real origin of the effect unexplained. Therefore development
of new methods of data analysis (the $z$-scaling is one of them)
 to clarify features of multi-particle production is of interest.

The energy and multiplicity independencies
of scaling function $\psi(z)$ for charged hadrons can be used to study
 the dependence of  $<p_T>$
for non-identified charged on $dN_{\rm ch}/d\eta$ and $\sqrt s$.
The results obtained in previous section allow us to construct
transverse momentum distributions $dN/dp_T$ over a wide kinematical range
for identified hadrons and calculate mean $p_T$ as follows
\begin{equation}
<p_T> =
{
{\int_{p_T^{\rm min}}^{\infty} p_T (dN/dp_T) dp_T } \over
 { \int_{p_T^{\rm min}}^{\infty}(dN/dp_T) dp_T}
}.
\label{eq:r25}
\end{equation}
Here $ p_T^{min}$ is the minimal transverse momentum of detected particle.
Figure 4 demonstrates
dependence of $<p_T>$ for produced charged hadrons
on multiplicity density $dN_{\rm ch}/d\eta $ (a) and
collision energy $\sqrt s $ (b).
As seen from Fig. 4 multiplicity and energy dependencies of $<p_T>$ reveal
monotonous growth with  $dN_{\rm ch}/d\eta$.
Similar behavior of $<p_T>$ were found for strange particles
$(K_S^0, \Lambda)$ produced in $pp$ collisions at RHIC energies \cite{Witt}.
We observe no indications on jump or sharp rise of $<p_T>$
up to highest values of multiplicity density.
Experimental verification of the predictions is of interest
for searching for phase transition of nuclear matter.

{\subsection{Jets}}

The STAR collaboration published  new data \cite{STAR_jet}
on cross section of jet production in $pp$ collisions at RHIC.
The data cover the kinematical range of pseudorapidity
$0.2<\eta<0.8$ and transverse momentum  $p_T = (5-50)$~GeV/c.

\begin{figure}
\hspace*{0mm}
\includegraphics[width=70mm,height=70mm]{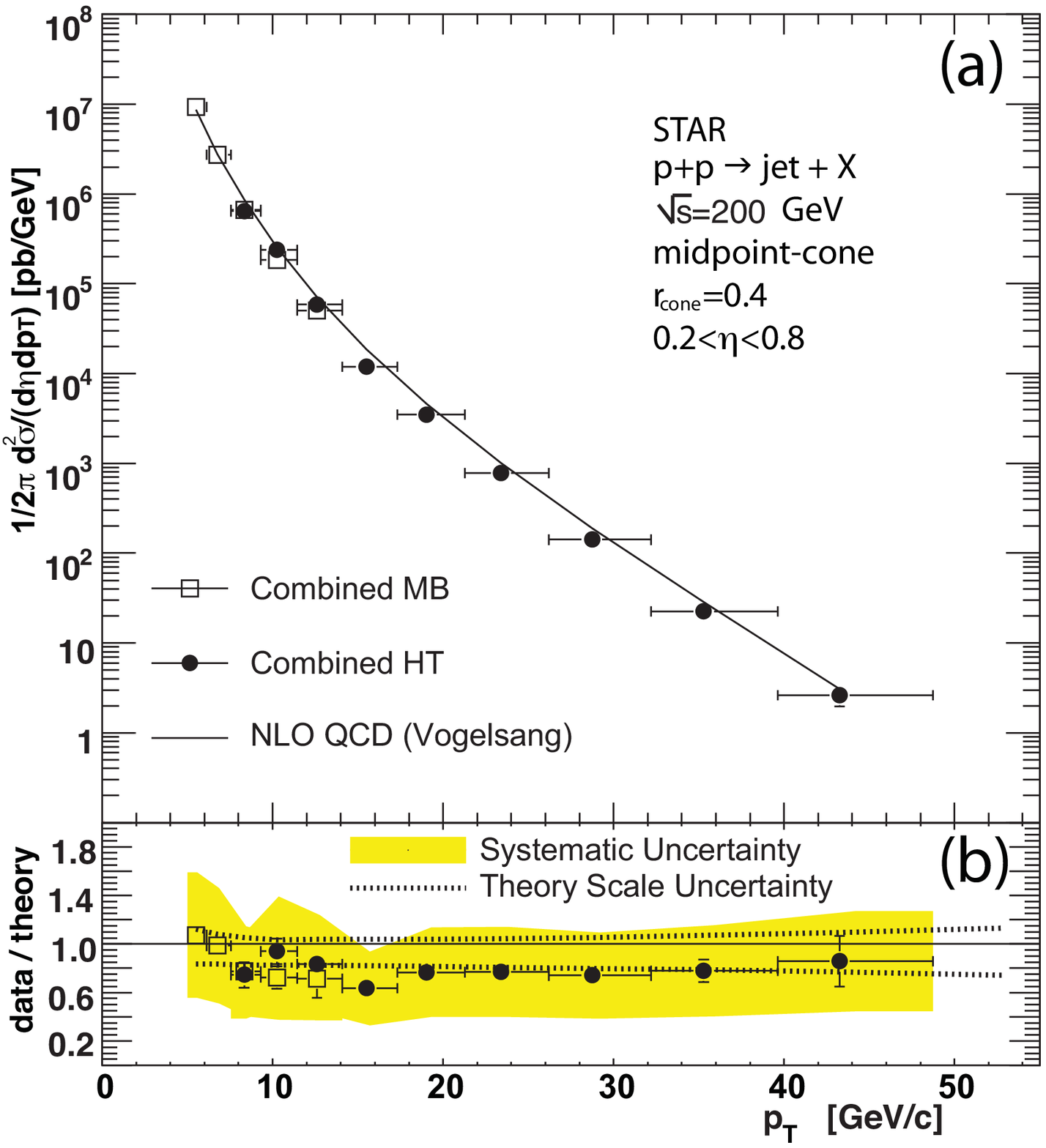}
\hspace*{10mm}
\includegraphics[width=70mm,height=70mm]{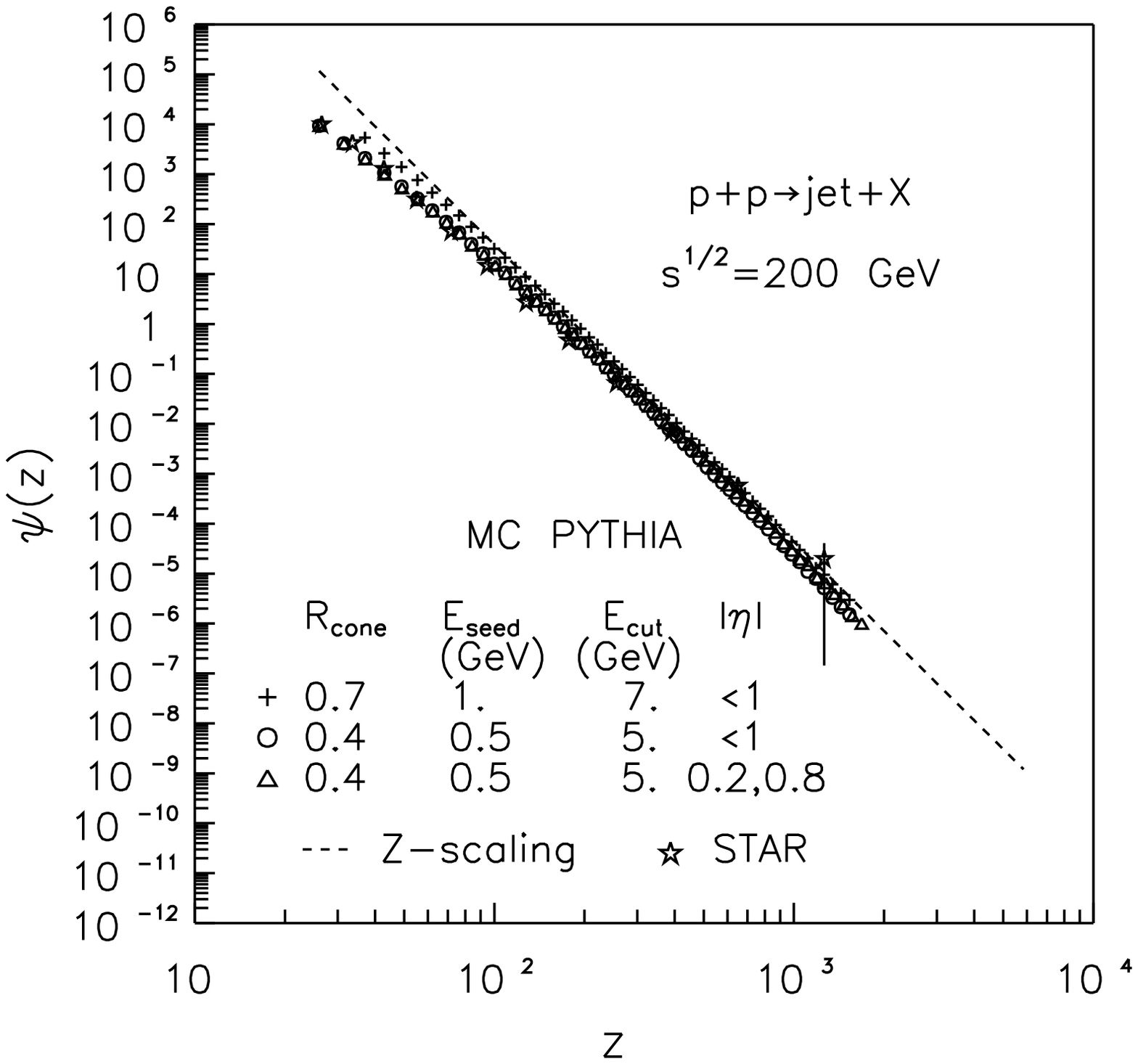}
\hspace*{40mm} (a) \hspace*{80mm} (b)
\caption{Transverse spectra of jet production in $pp$ collisions
at $\sqrt s =200$~GeV and $0.2< \eta <0.8$ in $p_T$ (a) and $z$ (b) presentation.
Experimental data obtained by STAR collaboration are taken from \cite{STAR_jet}.
MC results and $z$-scaling prediction are shown by points ($+, \circ, \triangle$)
and the dashed line, respectively.
}
\end{figure}

Figure 5(a) shows inclusive differential cross section
for the $p+p\rightarrow jet+X$ process at $\sqrt s =200$~GeV
measured by the STAR collaboration.
The NLO QCD calculated results
with the CTEQ6M parton distribution functions
at equal factorization and renormalization scales $\mu_R=\mu_F=p_T$
demonstrate satisfactory agreement with the data.
Comparison of the STAR data with MC results \cite{Ded_Tok}
and predictions of $z$-scaling in $z$  presentation is shown in Fig. 5(b).
The sensitivity of the scaling function to the choice of the parameters
$E_{seed}, R, E_{cut}$ of the jet-finding algorithm is observed.
It enhances as transverse momentum of jet decreases.
The dependence of ${\eta}^{jet}$ on $p_T^{jet}$ was used for construction
of the scaling function.
Note that the shape of the scaling function $\psi(z)$
for $p_T< 10$~GeV/c can not be described by the power law
$\psi(z)\sim z^{-\beta}$.
We see that both MC simulation results and STAR data are in a good agreement
with $z$-scaling predictions for $p_T^{Jet}>25$~GeV/c $(z>180)$.
The value of the slope parameter $\beta $ is found to be $\beta =6.01 \pm 0.06$ for
Monte Carlo results at $\sqrt s = 200$~GeV.
It is  compatible with $\beta=5.95 \pm
0.21$ obtained at lower energies $\sqrt s = 38.8, 45, 63$~GeV \cite{Zjet}.
For precise test of the asymptotic behavior of $psi(z)$ predicted by the $z$-scaling
and for verification of the property in the framework of QCD
measurement of jet spectra in $pp$ collisions at RHIC energies
with higher accuracy for high $p_T$ is necessary.

{\subsection{Self-similarity of charged particle production in $AA$}}

In this section we study the multiplicity dependence of $p_T$
and $z$ presentations of the experimental data \cite{STAR200,PHOBOS_Cu}
on inclusive cross sections
of charged hadrons produced in heavy ion collisions at RHIC.


The important ingredient of $z$-scaling is the multiplicity
density $dN_{ch}/d\eta(s,\eta)$ as a function
of the collision energy $\sqrt s $ and pseudorapidity $\eta$.
The scaling variable $z$
is proportional to $[dN_{ch}/d\eta(s,\eta)]^{-1}$ at $\eta=0$
while the function $\psi(z)$ is expressed via multiplicity density
depending on the energy $\sqrt  s$ and pseudorapidity $\eta$.
Using the special selection of events the transverse hadron spectra
were measured by the E735 collaboration
at highest $dN_{ch}/d\eta|_{0} \simeq 26$ \cite{mult_E735}.
The strong sensitivity of spectra to $dN_{ch}/d\eta$  was observed
to increase  with $p_T$.
The difference between cross sections corresponding  highest and lowest
multiplicity density was found to be about order of magnitude.
The value of multiplicity density of selected events for $\bar{p}p$ collisions
was larger than $dN_{ch}/d\eta|_{0}/(0.5N_p)$
 measured in central nucleus-nucleus collisions
at  AGS, S$\rm \bar p$pS and RHIC.
The multiplicity density in central $PbPb$ collisions
at $\eta=0$ at LHC energies
is expected to be about 8000 particles per unit of
pseudorapidity.
The regime of particle production at very high multiplicity
density is believed to be more preferable for searching
for clear signature of QGP formation.
\begin{figure}
\hspace*{45mm}
\includegraphics[width=85mm,height=85mm]{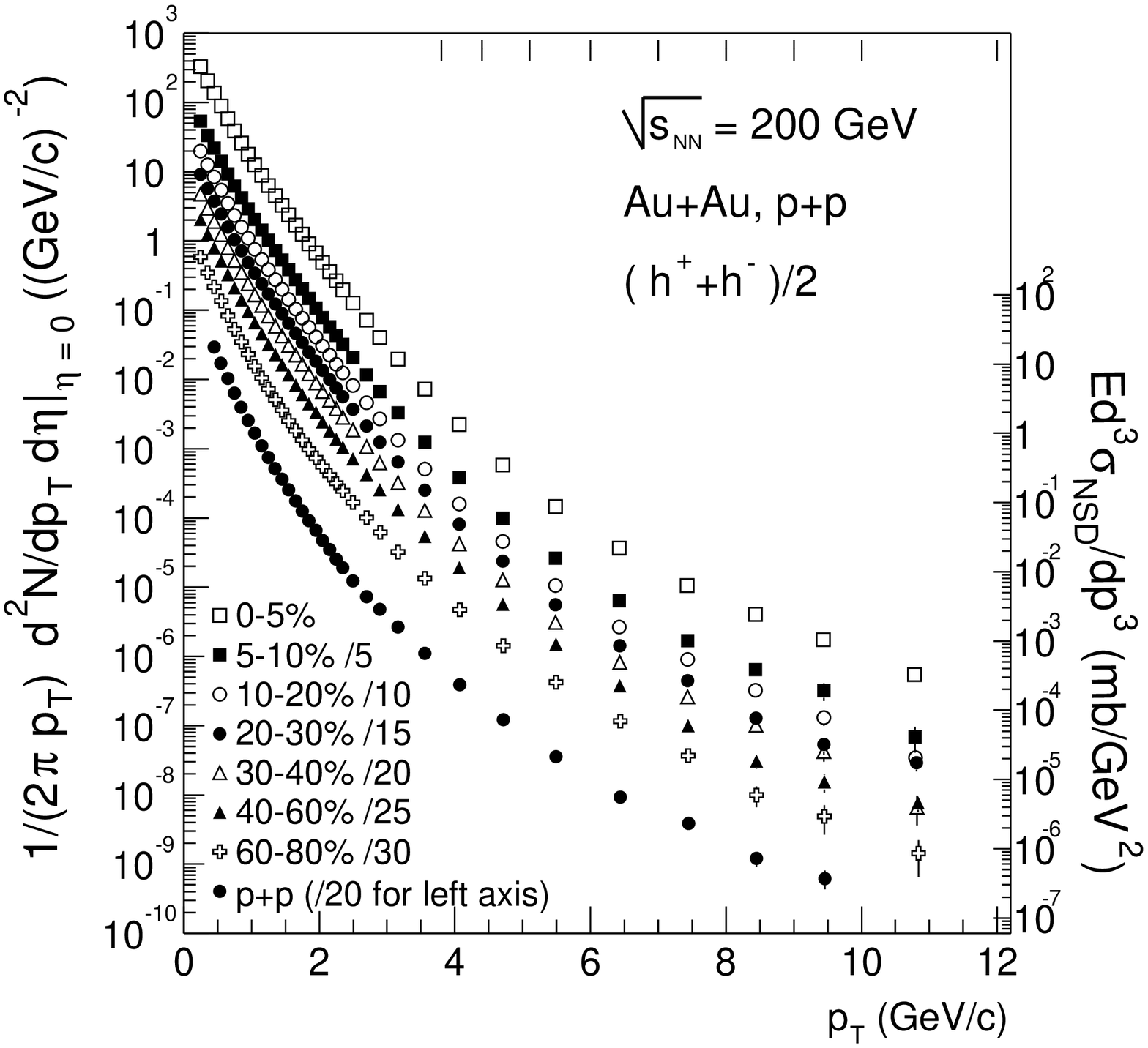}
\vskip 5mm
\caption{Transverse momentum spectra of charged hadrons produced in $AuAu$ and $pp$
collisions at RHIC  as a function of multiplicity density.
Experimental data obtained by STAR Collaboration are taken from \cite{STAR200}.}

\vskip 20mm

\hspace*{0mm}
\includegraphics[width=70mm,height=70mm]{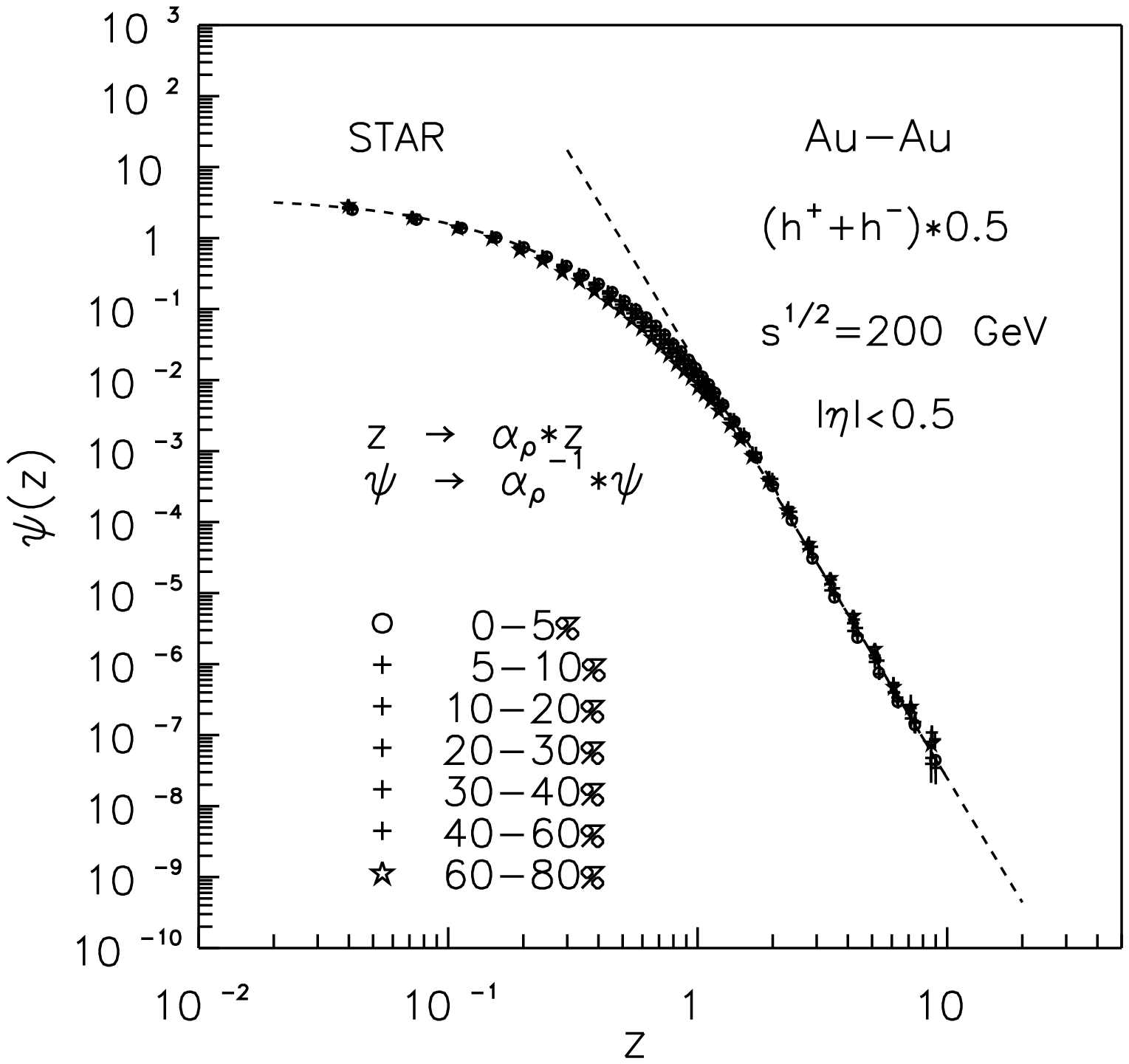}
\hspace*{10mm}
\includegraphics[width=70mm,height=70mm]{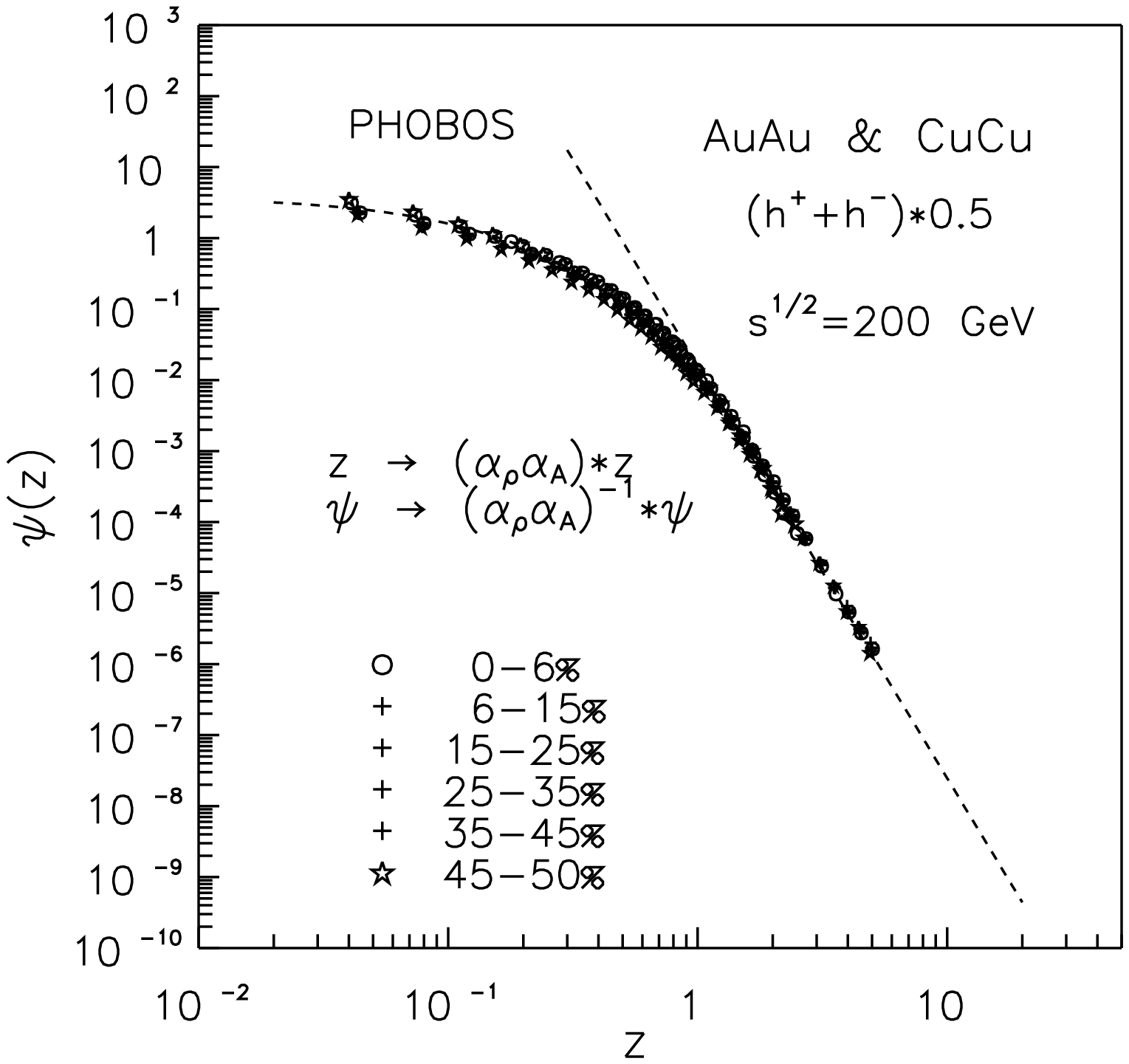}
\hspace*{40mm} (a) \hspace*{80mm} (b)
\caption{Transverse momentum spectra of charged hadrons produced in $AuAu$ and $CuCu$
collisions at RHIC in $z$ presentation as a function of multiplicity density.
Experimental data are taken from \cite{STAR200,PHOBOS_Cu}.}
\end{figure}

Figure 5 shows the dependence of the spectra of
charged hadron production in $AuAu$ collisions
on the transverse momentum $p_T$  at the energy $\sqrt s = 200$~GeV
and over the  pseudorapidity
range $|\eta|<0.5$ for different centralities \cite{STAR200}.
The data cover a wide transverse momentum range, $p_T = 0.2-11$~GeV/c.
A strong sensitivity of high-$p_T$ spectra to multiplicity density
$dN_{ch}/d\eta$ is observed.
We use values of the fractal dimensions
$\delta_N=0.7$ and $\epsilon =0.7$ to study the dependence of transverse spectra
on multiplicity density $dN_{ch}/d\eta$ in $AuAu$ and $CuCu$ collisions.
The scaling behavior of the data (Fig.6(a)) are restored
at $c=0.25$ and under the simultaneous transformation
of the variable $z$ and function $\psi(z)$:
 $z \rightarrow \alpha_{\rho}z$ and $\psi \rightarrow \alpha_{\rho}^{-1} \psi$
($\rho \equiv dN_{ch}/d\eta$).
The power behavior (the straight dashed line in Fig.6(a))
 of the scaling function, $\psi(z) \sim z^{-\beta}$,
for high $z$ is observed. The soft regime
of particle production demonstrates self-similarity
 in $z$  presentation for low $z$ as well.

Figure 6(b) demonstrates $z$ presentation of the transverse spectra
of charged hadrons produced in $AuAu$ and $CuCu$ collisions
in the central rapidity range
at $\sqrt s =200$~GeV as a function of multiplicity density.
The data obtained by the PHOBOS collaboration \cite{PHOBOS_Cu}
reach transverse momenta up to 4~GeV/c.
The dashed lines are obtained by
fitting of $\psi(z)$  corresponding to the STAR data \cite{STAR200}
at $\sqrt s =200$~GeV.
For comparison  scaling functions for different nuclei the  transformation
of $z \rightarrow \alpha_{A}z$ and $\psi \rightarrow \alpha_{A}^{-1} \psi$
($\rho \equiv dN_{ch}/d\eta$) was used. The scaling was found to be
 restored at the same value of $c=0.25$.
The $z$  presentation demonstrates a good compatibility
of the PHOBOS and STAR data over a kinematical range measured
by the PHOBOS collaboration \cite{PHOBOS_Cu}.

Thus based on the obtained  results we conclude
that mechanism of charged hadron production in $AuAu$ and $CuCu$
collisions at $\sqrt s =200$~GeV
reveals self-similarity and fractality over a wide
range of transverse momentum and multiplicity density.

{\section{$z$-Scaling in $\bar{p}p$ collisions at Tevatron}}

In this section we present results of our analysis
of new data obtained by the D0 and CDF Collaborations at Tevatron in Run II.
We verify properties of $z$-scaling established in previous papers
such as the energy and angular independence of the scaling function $\psi(z)$
for particle (hadrons, direct photons, jets) production
at high $p_T$. The hypothesis of flavor independence of $\psi(z)$
for high $z$ is used for prediction of spectra of different particles
($J/\psi,\Upsilon, D^0,B^+,Z,W^+$) as well.

{\subsection{Direct photons}}

Recently the D0 collaboration published the new data \cite{D0_g}
on inclusive cross sections
of direct photons produced in $\bar{p}p$ collisions at $\sqrt s=1960$~GeV.
The data cover the momentum  $p_T=30-250$~GeV/c
and pseudorapidity $|\eta|<0.9$ range.
The data together with data obtained by D0 in Run I are presented in Fig. 7(a).
The strong angular dependence of cross section is observed.
It increases with $p_T$ and reaches about order of magnitude at $p_T=100$~GeV/c.

\begin{figure}
\hspace*{10mm}
\includegraphics[width=60mm,height=60mm]{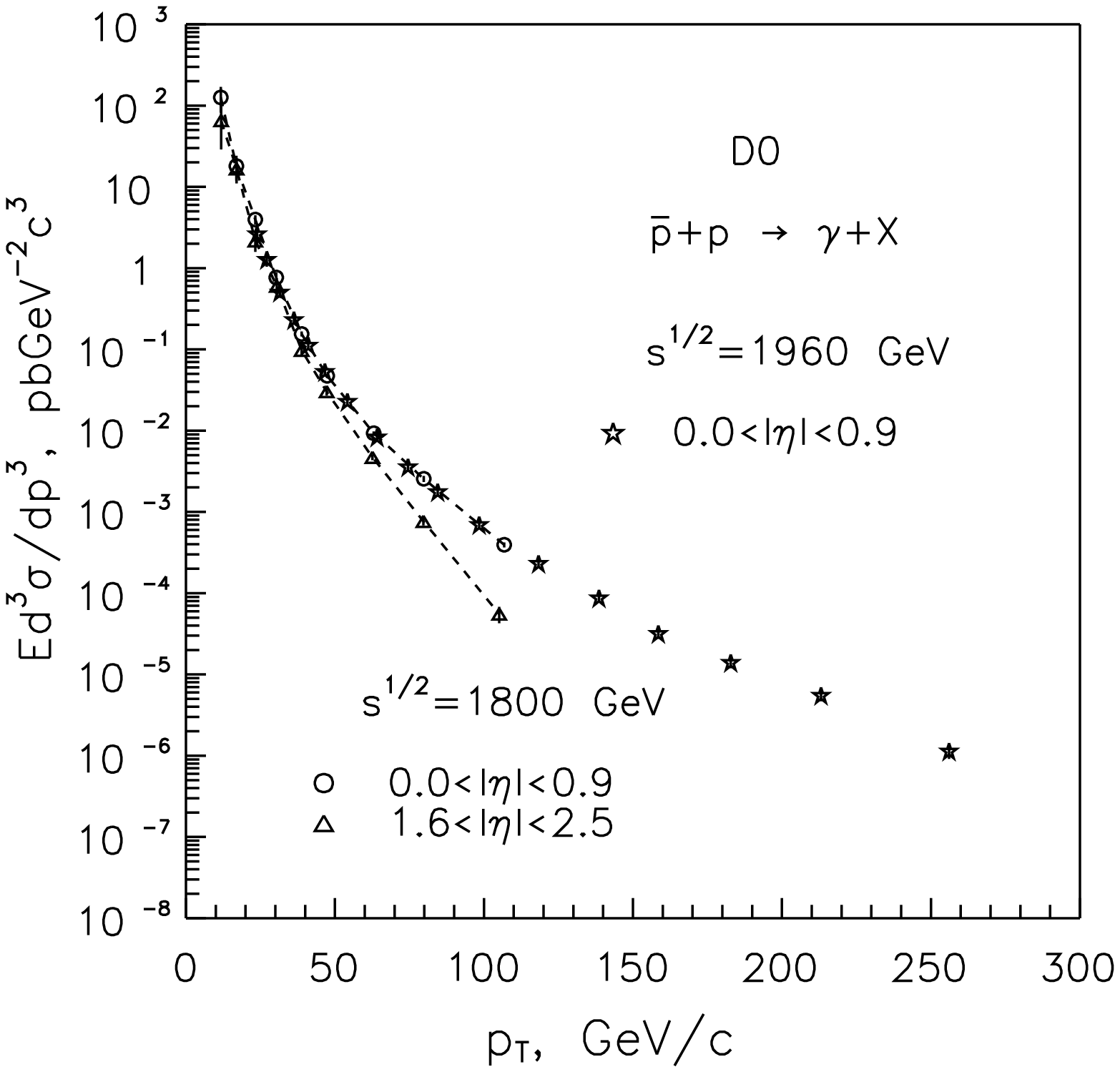}
\hspace*{15mm}
\includegraphics[width=60mm,height=60mm]{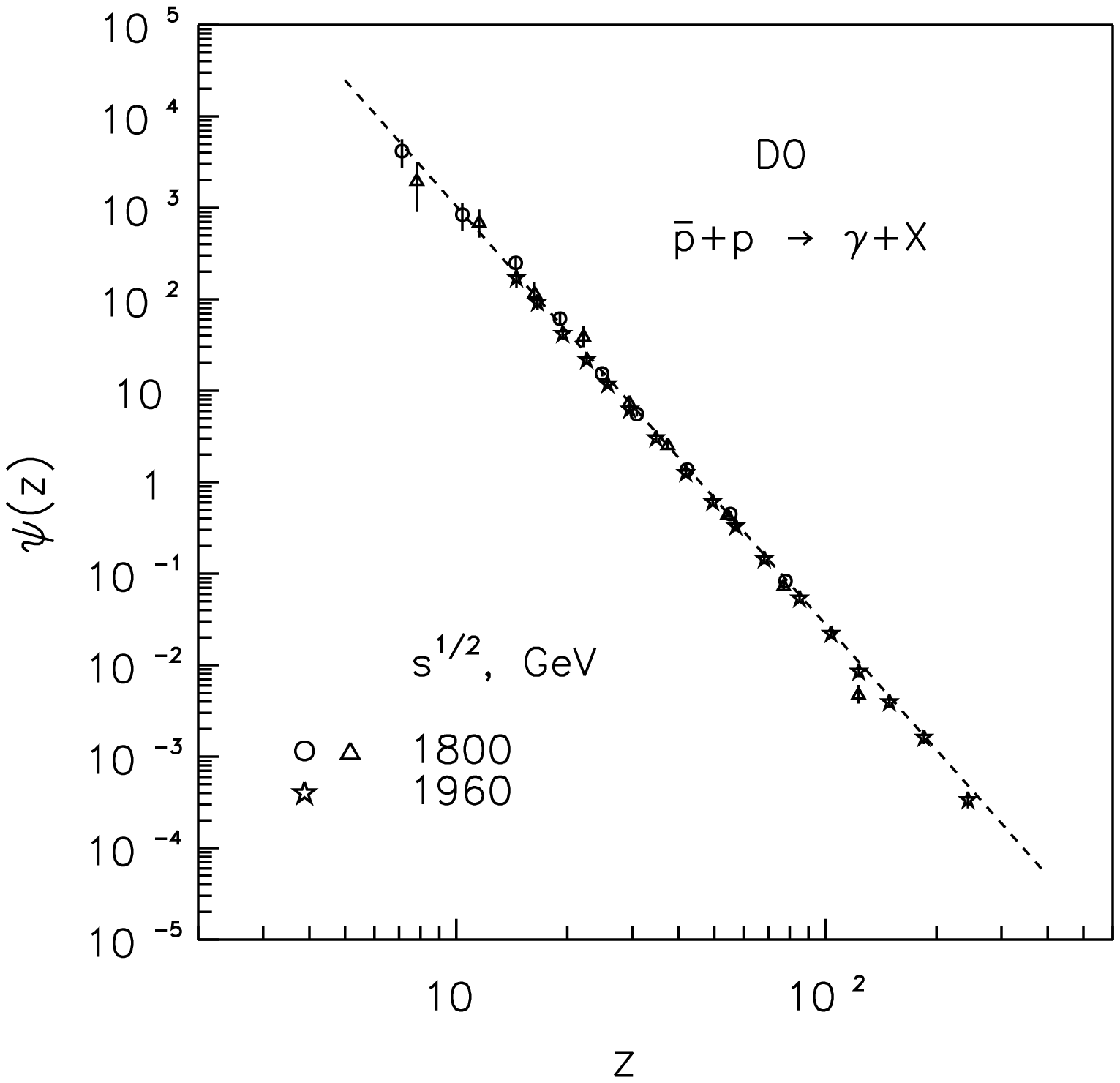}
\hspace*{40mm} (a) \hspace*{80mm} (b)
\caption{ Spectra of direct photons produced in $\bar{p}p$ collisions
in $p_T$ and $z$ presentations. Experimental data obtained
by D0 Collaboration in Run I and II are taken from \cite{D0_g}.}
\end{figure}
The $z$  presentation of the same data is shown in Fig. 7(b).
One can see that new experimental data confirm features
(the energy and angular independence of $\psi(z)$)
of $z$-scaling for direct photon production in $\bar {p}p$
and $pp$ collisions established in \cite{Zg,Zg1}.
The power law, $\psi(z)\sim z^{-\beta}$, is observed over a wide range of $z$.
We consider the independence of the slope parameter $\beta $ on kinematical variables
$(\sqrt s, p_T$ and $\eta)$ as an evidence of self-similarity and fractality
of photon production.
It can mean that structure of a photon itself
at small scales looks like structure of other particles (hadrons)
characterized by fractal dimension(s)~{\footnote{In paper \cite{Zg}
the fractal dimensions
$\delta_1, \delta_2$ are introduced as parameters for description
of the fractal  measure $z=z_0\Omega^{-1}$,
$\Omega=(1-x_1)^{\delta_1}(1-x_2)^{\delta_2}$}}.

{\subsection{Jets}}

In this section we present results of analysis of new data
on inclusive cross sections
of jet production in $\bar{p}p$ collisions at $\sqrt s = 1960$~GeV
obtained by the D0 and CDF collaborations at Tevatron \cite{D0_CDF_jet}
and compare them with our previous results \cite{Zjet}.

\begin{figure}
\hspace*{10mm}
\includegraphics[width=60mm,height=60mm]{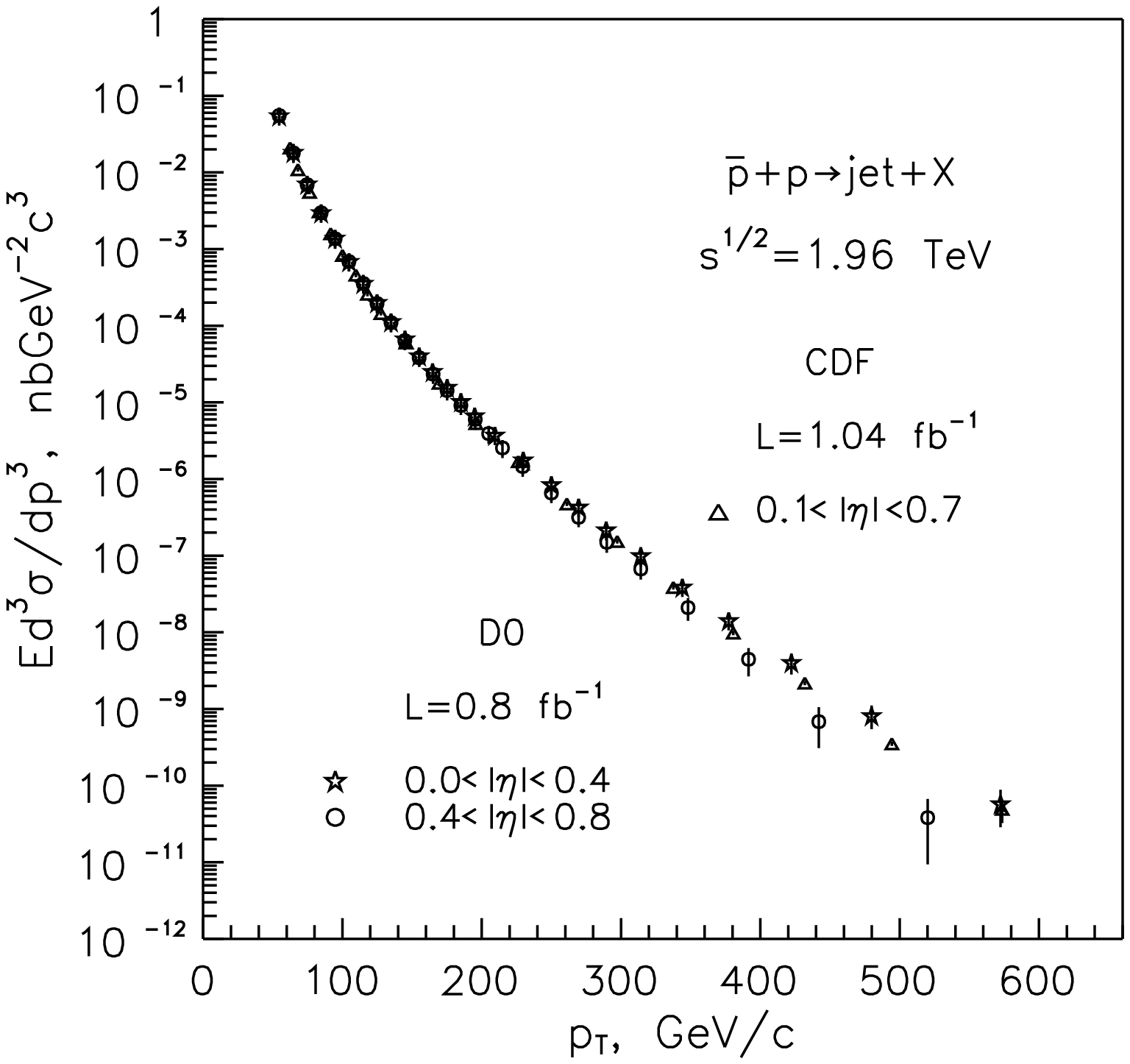}
\hspace*{15mm}
\includegraphics[width=60mm,height=60mm]{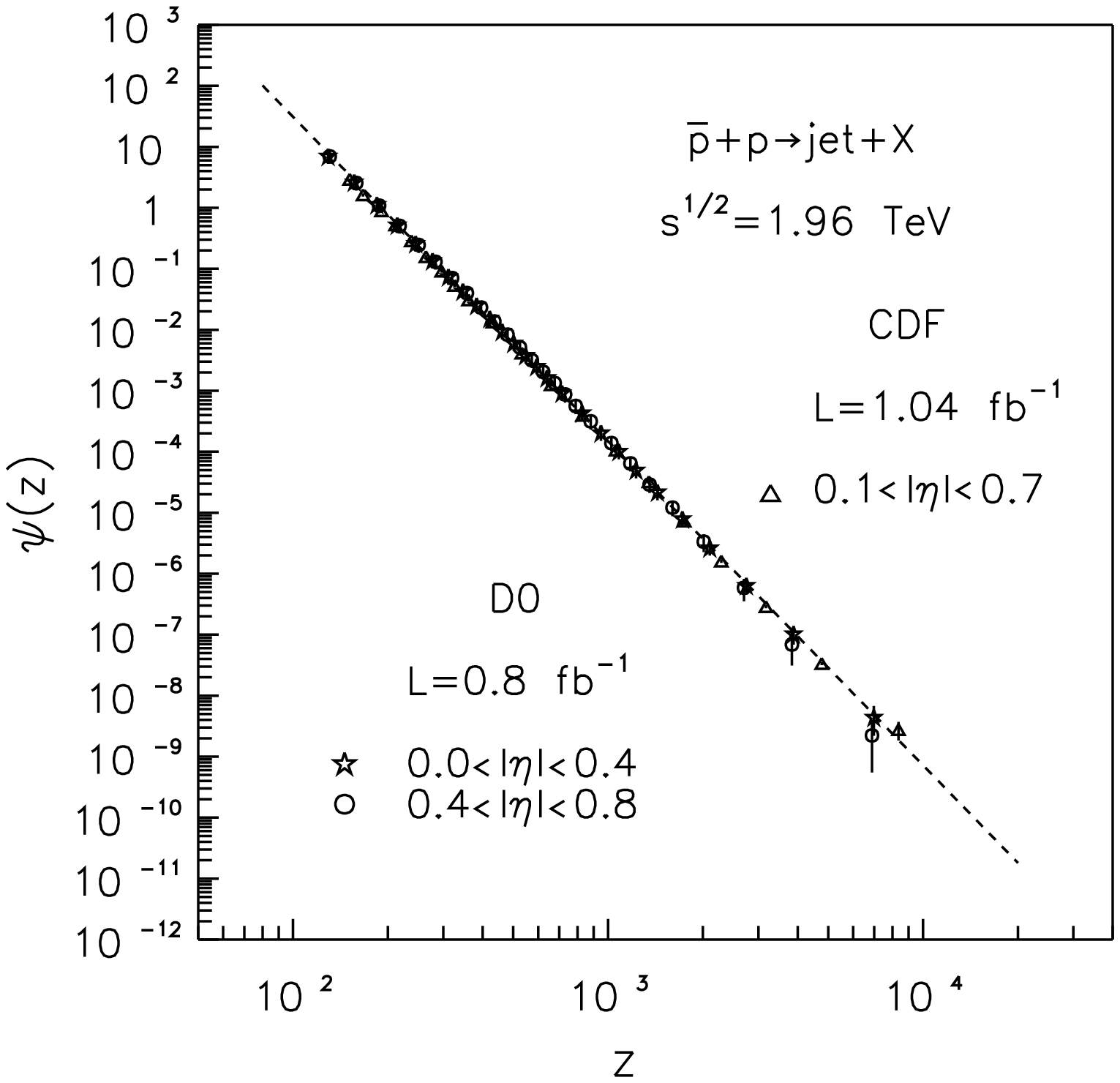}
\hspace*{45mm} (a) \hspace*{75mm} (b)
\caption{Spectra of jet production in $\bar{p}p$ collisions
in $p_T$ and $z$ presentations.
Experimental data obtained by the D0 and CDF collaborations
are taken from \cite{D0_CDF_jet}.}
\end{figure}


\begin{figure}
\hspace*{10mm}
\includegraphics[width=60mm,height=60mm]{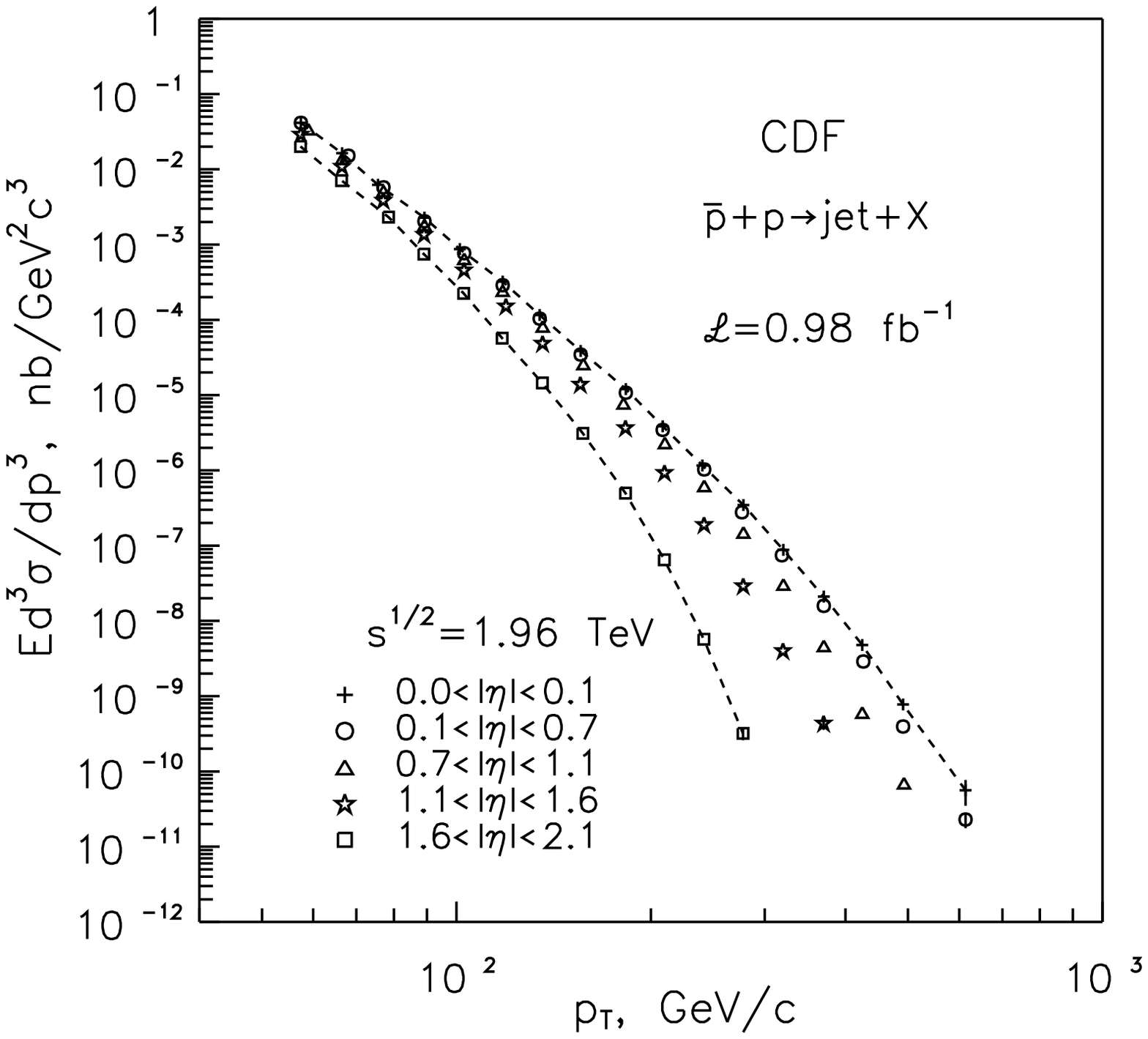}
\hspace*{15mm}
\includegraphics[width=60mm,height=60mm]{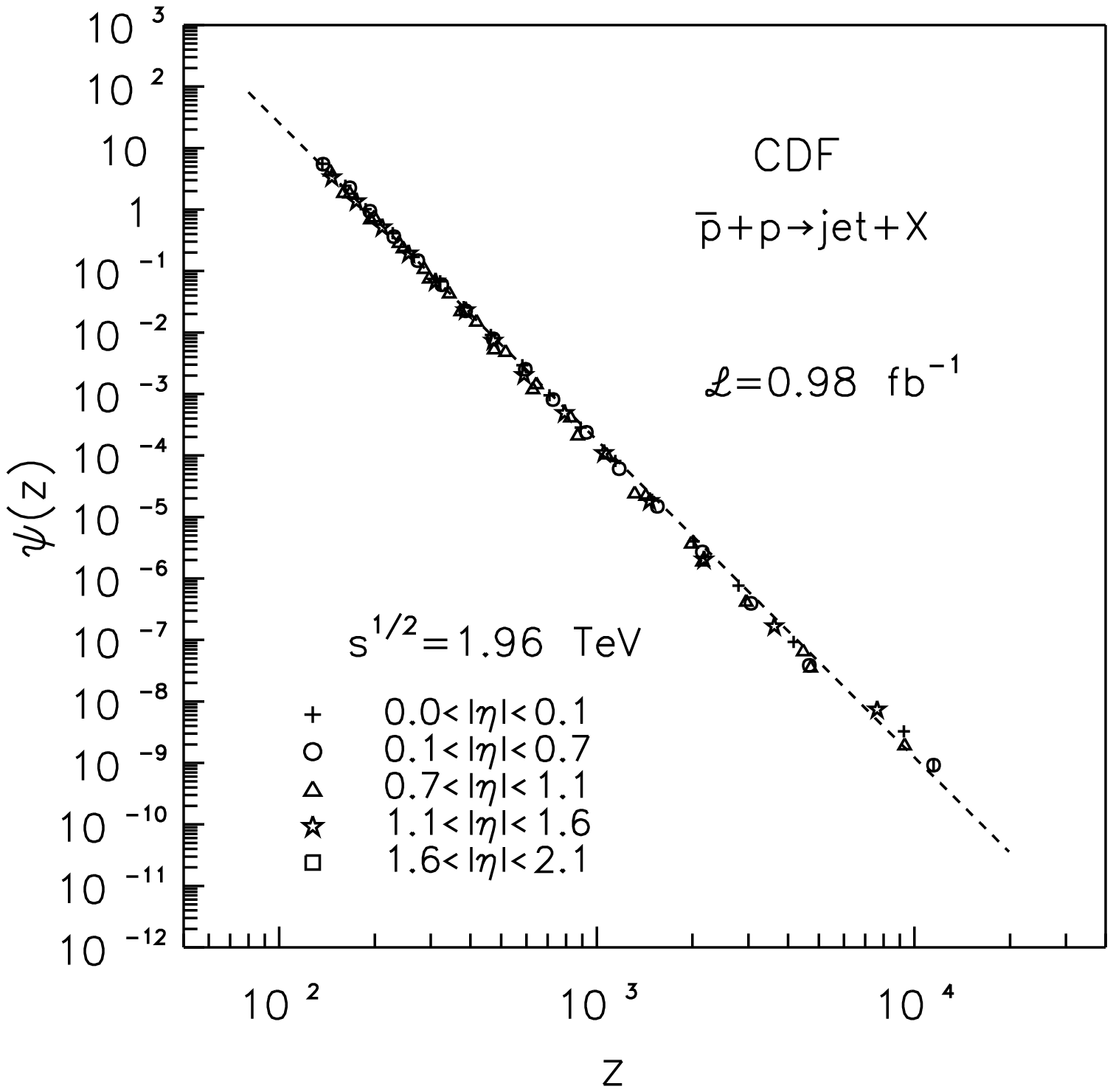}
\hspace*{45mm} (a) \hspace*{75mm} (b)
\caption{ Angular dependence of spectra of jet
production in $\bar{p}p$ collisions
in $p_T$ and $z$ presentations.
Experimental data are taken from \cite{D0_CDF_jet}.}
\end{figure}

Production of hadron jets at Tevatron probes the highest
momentum transfer region currently accessible  and thus
potentially sensitive  to a wide  variety of new physics.
The information on inclusive jet cross sections at high
transverse momentum range is the basis to test QCD, to extract
the parton distribution functions and to constrain
uncertainties for gluon distribution in the high-$x$ range.
In Run II, as  mention in \cite{Matthias}, the measurement of jet
production  and the sensitivity to new physics will profit
from the large integrated luminosity  and the higher cross
section, which is associated with  the increase in the center-of-mass
energy from 1800 to 1960~GeV.
Therefore the test of $z$-scaling for jet production in
$\bar pp$ collisions in new kinematic range is of great interest
to verify scaling features established in our previous analysis \cite{Zjet}.

Figure 9(a) shows the new Run II data \cite{D0_CDF_jet}
on the inclusive jet cross section at $ \sqrt s =1960$~GeV.
The pseudorapidity range covered by the D0 and CDF collaborations
corresponds to $|\eta|<0.8$ and $0.1<|\eta|<0.7$, respectively.
The transverse momentum of jet changes from 50 up to 560~GeV.
The $z$  presentation of data  is shown in Fig.9(b).
As seen from Fig.9(b) the D0 and CDF data are compatible each other.
The energy independence of $\psi(z)$ is observed up to $z \simeq 4000$.
Asymptotic behavior of the scaling function is described by the power
law, $\psi(z)\sim z^{-\beta}$ (the dashed line in Fig.9(b)).
The slope parameter $\beta$ is energy independent over a wide $p_T$-range.
Note that results of present analysis of new D0 data are
in a good agreement with our results \cite{Zjet} based on the data
obtained by the same collaboration in Run I.
The energy independence and the power law (the dashed line in Fig.9(b))
of the scaling function $\psi(z)$ are found to be as well.

The angular dependence of inclusive cross section of jet production
in $\bar{p}p$ collisions at $\sqrt s =1960$~GeV was investigated
by the CDF collaboration \cite{D0_CDF_jet}.
The experimental data cover the rapidity range $|\eta|<2.1$.
The highest transverse energy carried by one jet was determined to be ~600~GeV.
As seen from Fig. 10(a) the transverse spectra demonstrate
the strong dependence on pseudorapidity of produced jet.
The $z$  presentation of the same data is shown in Fig. 10(b).
It demonstrates the angular independence and the power behavior
of $\psi(z)$. We would like to emphasize that these results
are new confirmation of $z$-scaling. Jet production is usually considered as
signature of hard collisions of elementary constituents (quarks and gluons).
Therefore the obtained result means that interaction of constituents,
their substructure
and mechanism of jet formation reveal properties
of self-similarity over a wide scale range (up to $10^{-4}$~Fm).

{\subsection{$b$-Jets}}

\begin{figure}
\hspace*{10mm}
\includegraphics[width=60mm,height=60mm]{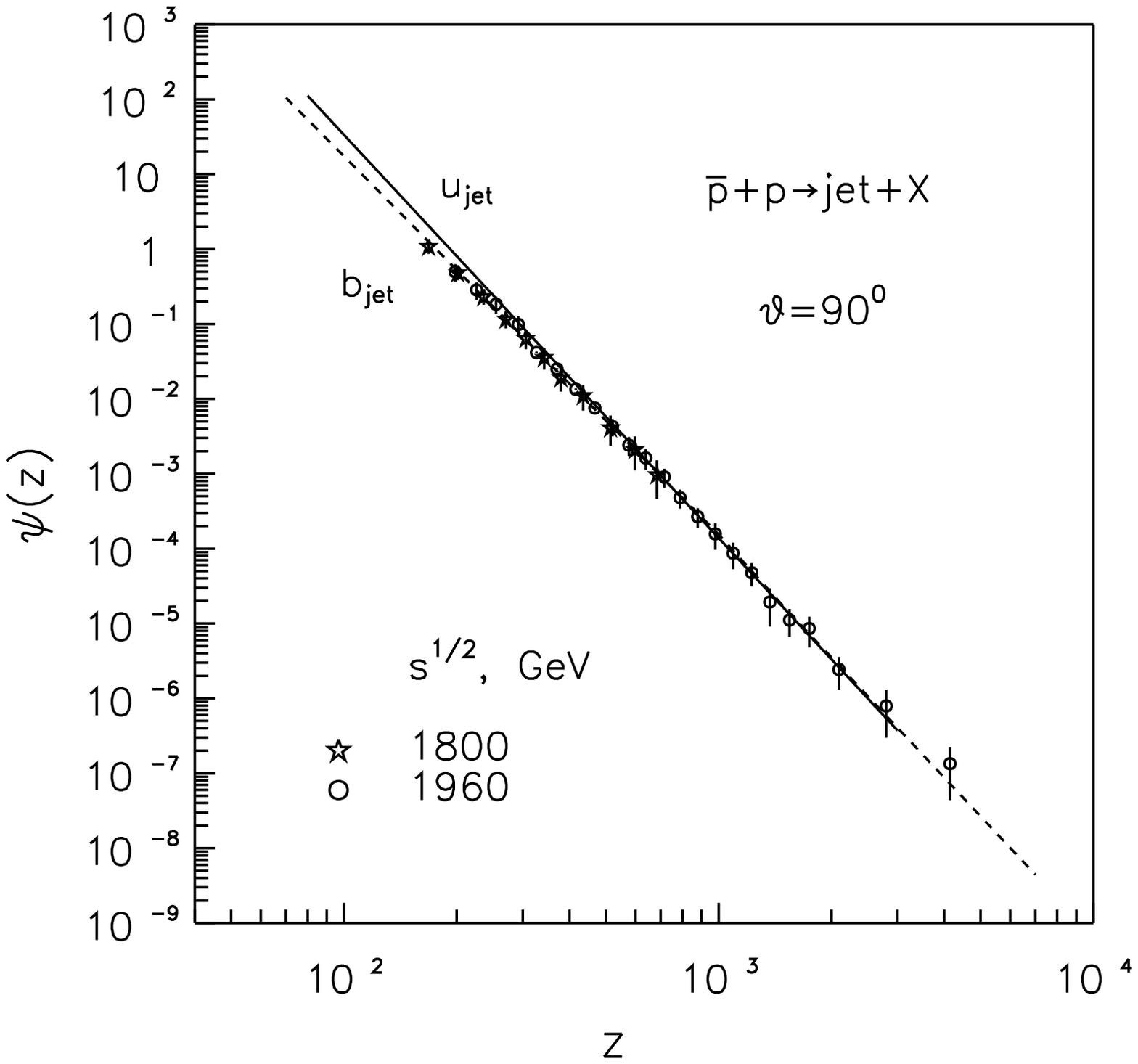}
\hspace*{15mm}
\includegraphics[width=60mm,height=60mm]{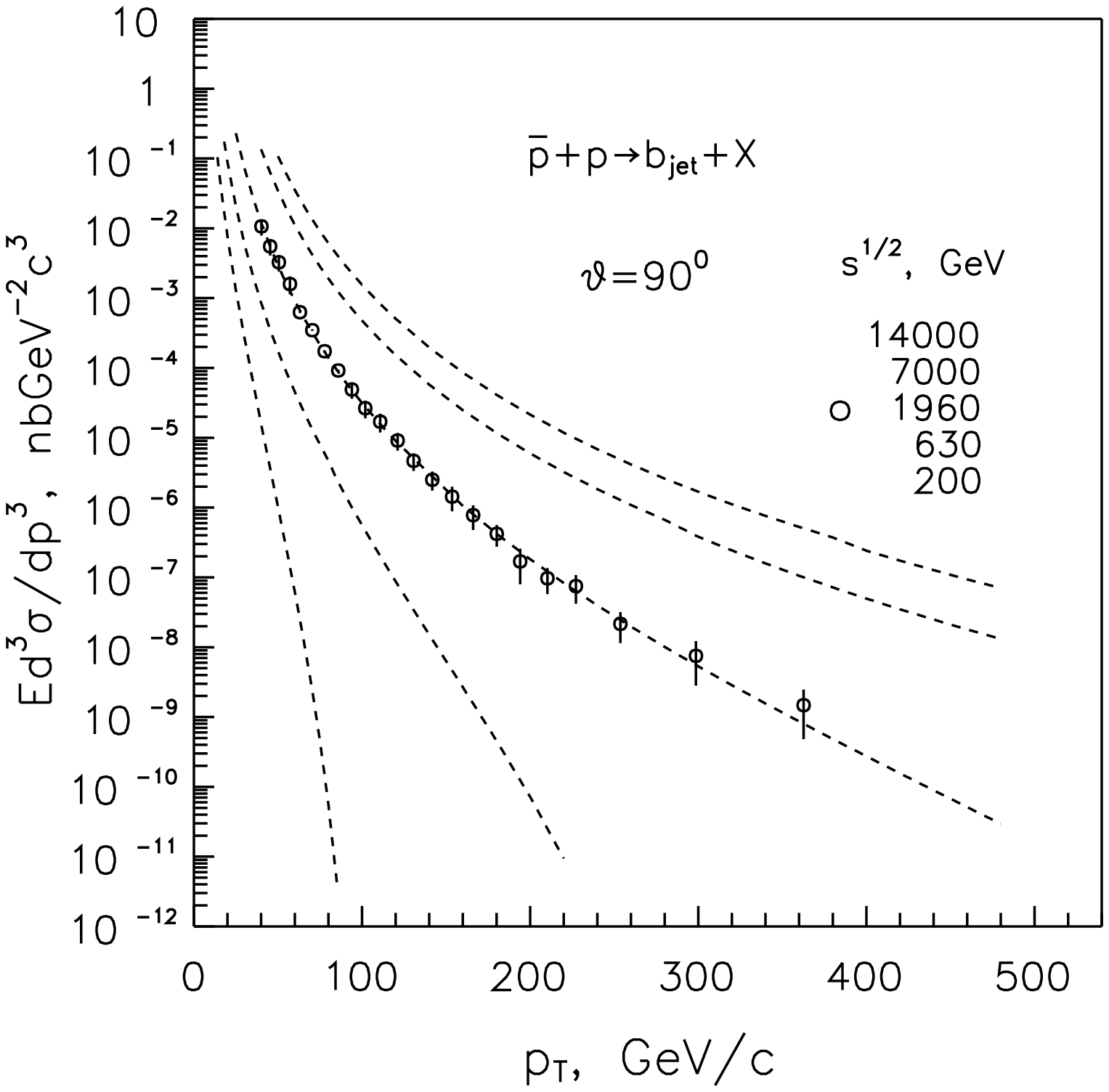}
\hspace*{45mm} (a) \hspace*{75mm} (b)
\caption{ Spectra of $b$-jets produced in $\bar{p}p$ collisions
in $z$ (a) and  $p_T$ (b) presentations.
Experimental data are taken from \cite{bjet}.}
\end{figure}

Flavor independence  of $z$-scaling means that value
of the slope parameter $\beta$ of the scaling function
$\psi(z)$ is the same for different types of produced hadrons.
The hypothesis is supported by results of analysis of hadron
($\pi^{\pm,0},K,\bar{p}$) spectra  for high $p_T$ in $pp$
and $p-A$ collisions \cite{Zflavor}.
The verification of the hypothesis is of interest
for understanding of mechanism of particle production
at very small scale. We assume that the transformation
of a point-like quark ($u,d,s,c,b,t$)
into real hadron produced at high $p_T$
is the self-similar process which is independent of the quark flavor.

The spectra of $b$-tagged jets
were measured \cite{bjet} by the D0 and CDF collaborations
at $\sqrt s =1800$ and 1960~GeV, respectively.
Transverse energy of jet was measured in the range $25-400$~GeV.
The data together with the CDF data obtained in Run I
on inclusive cross section for non-tagged ($u_{jet}$) jets
in $z$  presentation are shown in Fig. 11(a).
Both data demonstrate power behavior of $\psi(z)$ for high $z$.
Deviation  from the power law is observed for
$z<300$~{\footnote{We assume that for $s$- and $c$-tagged jets
the deviation will be smaller and for $t$-tagged jets larger
 than for $b$-tagged jets.}}.
The present calculation
was performed~{\footnote{The calculation procedure is described in \cite{Zjet}.}}
with the values of the fractal dimensions $\delta_1=\delta_2=1$.
We use the universality of asymptotic behavior of  $\psi(z)$ for non-tagged jets
and pre-asymptotic behavior of $\psi(z)$ for $b$-tagged jets for construction
of the scaling function  over a wide range of $z$.
Our predictions of inclusive cross sections of $b$-jet production
in $\bar{p}p$ collisions at $\sqrt s =200-14000$~GeV and
the CDF data ($\circ$) are shown in Fig. 11(b).

{\subsection{$J/\psi$ mesons}}

\begin{figure}
\hspace*{10mm}
\includegraphics[width=60mm,height=60mm]{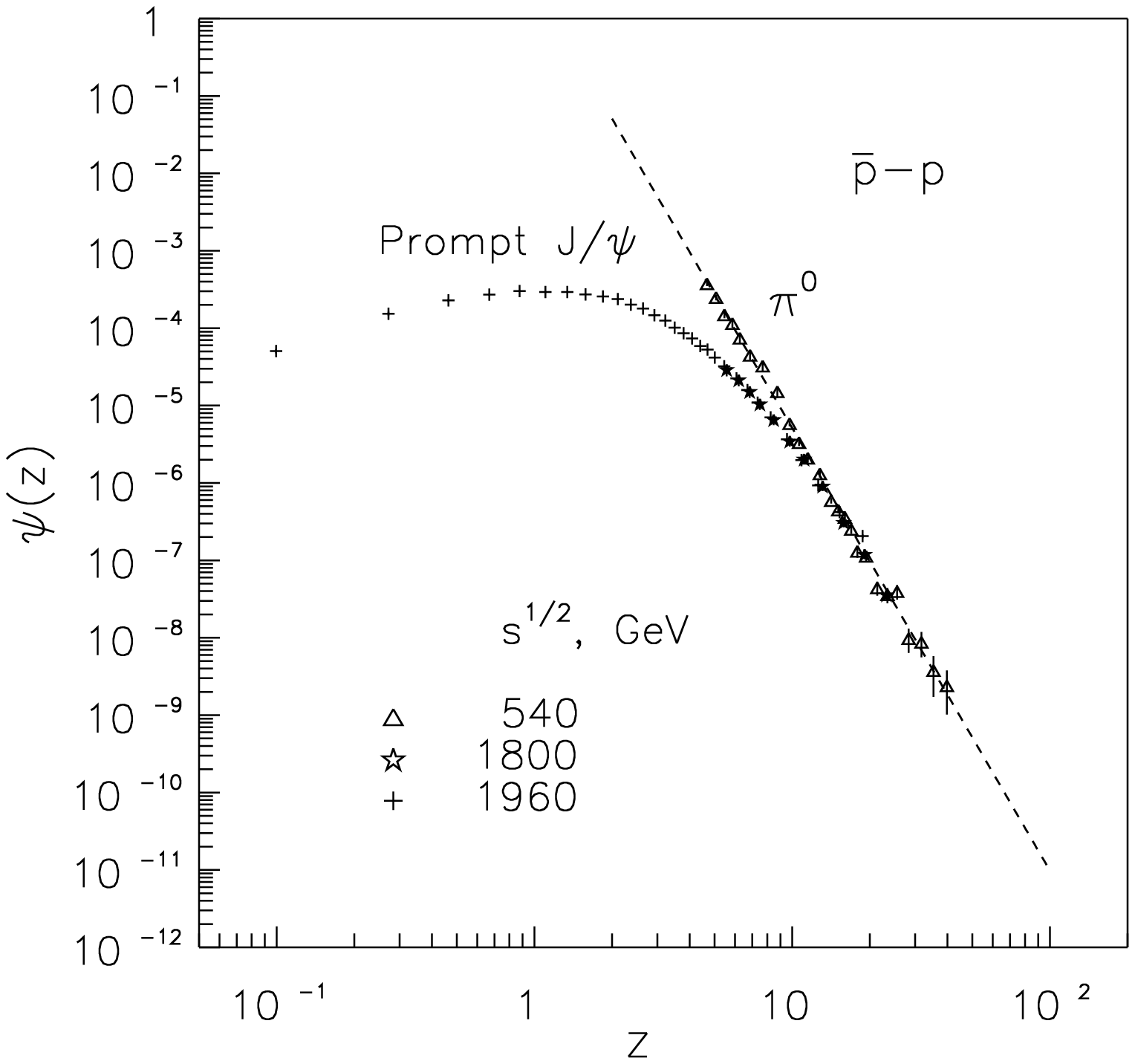}
\hspace*{15mm}
\includegraphics[width=60mm,height=60mm]{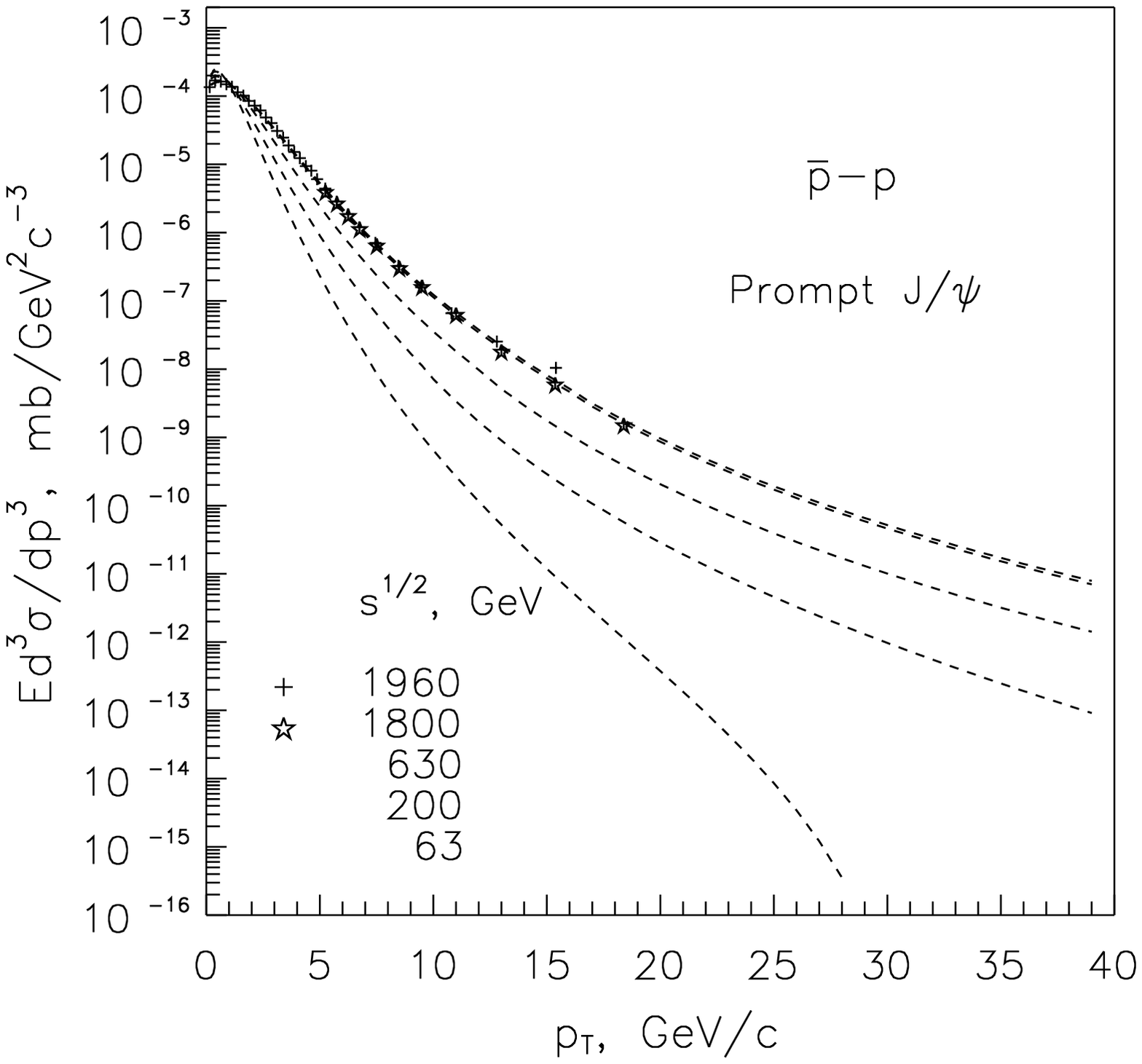}
\hspace*{40mm} (a) \hspace*{80mm} (b)
\caption{Spectra of $J/\psi$ mesons produced in $\bar{p}p$ collisions
in $z$ (a) and $p_T$ (b) presentations.
Experimental data are taken from \cite{Jpsi1,Jpsi2,Banner}.
}
\end{figure}

Here we present results of analysis \cite{Tok_Jpsi}
of the data \cite{Jpsi1,Jpsi2} on inclusive cross section of $J/\psi$
production in $\bar{p}p$ collisions.
The hypothesis of the flavor independence of $z$-scaling
was used for the analysis.
The experimental UA1 data \cite{Banner} on inclusive cross section
of $\pi^0$-mesons produced in $\bar{p}p$ collisions at $\sqrt s =540$~GeV
were used to construct the scaling function in the asymptotic region.
The scaling function of $\pi^0$ is shown by the dashed line in Fig.12(a).
It is described by the power law, $\psi(z)\sim z^{-\beta}$.
The value of the slope parameter $\beta$ was found to be $5.77\pm 0.02$
over a wide $p_T$ range.
To compare $z$  presentations for $J/\psi$ and $\pi^0$
the transformation of $z$ and $\psi$ in the form
$z\rightarrow \alpha_F\cdot z, \ \ \  \psi \rightarrow {\alpha_F}^{-1} \cdot \psi $
was applied.
The coefficient $\alpha_F$ was found to be a constant (2.33).
It was used to describe the overlapping region for both particles.
Note that the CDF Run I and Run II data demonstrate
a good matching in the region.
Strong deviation of scaling function for $J/\psi$ from the power law
is seen for $p_T<10$~GeV/c.
The energy independence of the scaling function was
used to predict transverse spectra of $J/\psi$ production
in $\bar{p}p$ collisions in the central rapidity range
at the energy $\sqrt s =63, 200, 630, 1800$ and 1960~GeV.
The calculated results and the CDF data \cite{Jpsi2}
are shown in Fig. 12(b) by the dashed lines and points, respectively.
One can see that the strong dependence
of inclusive cross section on $\sqrt s $ enhances with $p_T$.
Experimental test of the predicted results  is of interest
for understanding of mechanism of $J/\psi$ production
and properties of $z$-scaling.

{\subsection{$D^0$ mesons}}

Data on open charm production can provide understanding
of mechanism of particle formation depending
on the flavor quantum number
and test of QCD predictions. They could give additional
constraints on parton distribution and fragmentation
functions of charmed quark.
Such data allow us to study flavor dependent features
of $z$-scaling.

The inclusive charm meson cross sections in $\bar {p}p$ collisions
were measured
by the CDF collaboration in the central rapidity
$|y|<1$ and transverse momentum $p_T = 5-20$~GeV/c range
at $\sqrt s =1960$~GeV \cite{D0_meson}.
Here we present results of analysis in the framework of $z$-scaling
of the data corresponding to reconstructed decay mode $D^0\rightarrow K^-\pi^+$.
It was established that the prompt fraction
of $D^0$ meson production for each $p_T$ bin is $(86.6\pm0.4)\% $.
\begin{figure}
\hspace*{10mm}
\includegraphics[width=60mm,height=60mm]{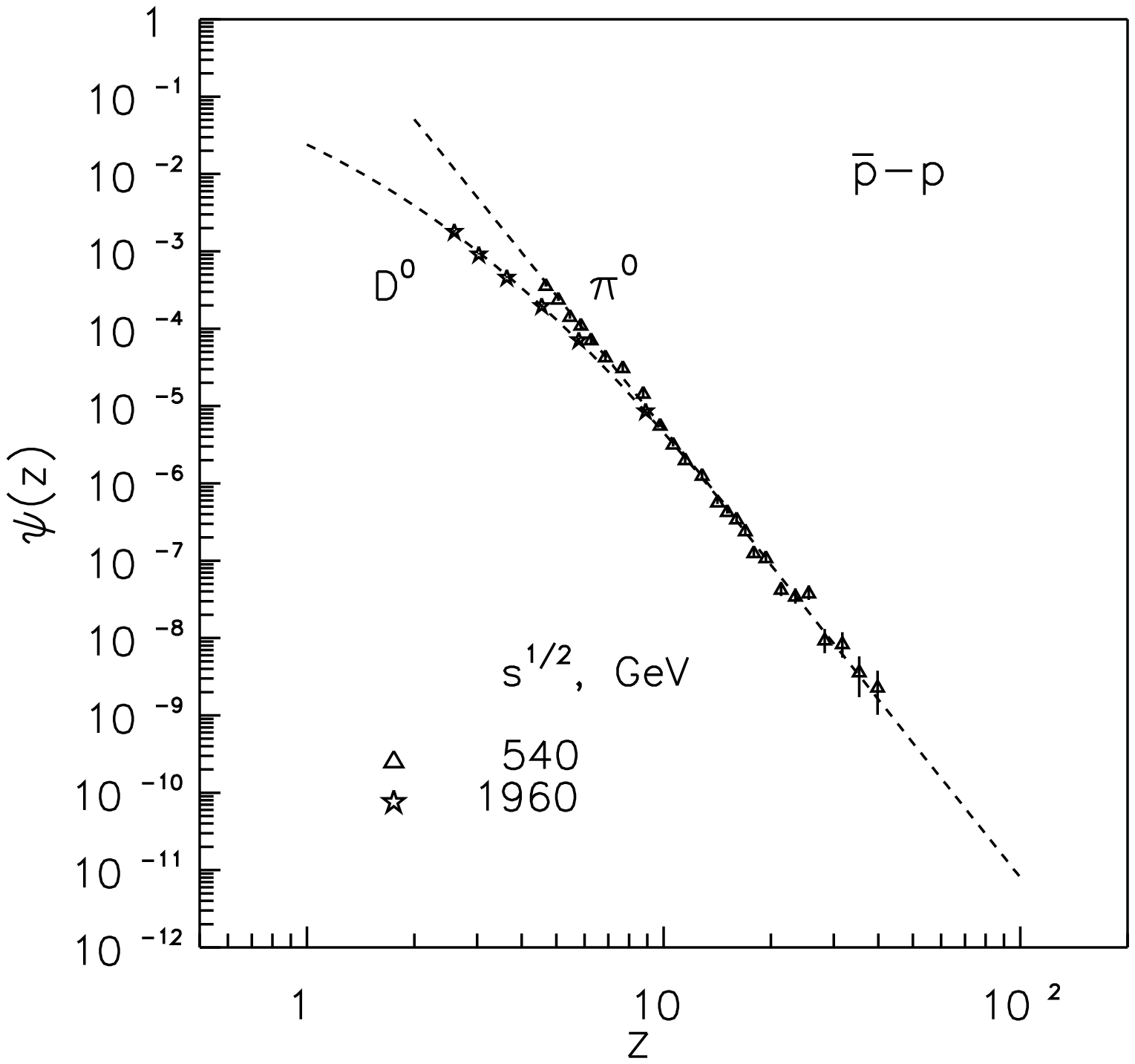}
\hspace*{15mm}
\includegraphics[width=60mm,height=60mm]{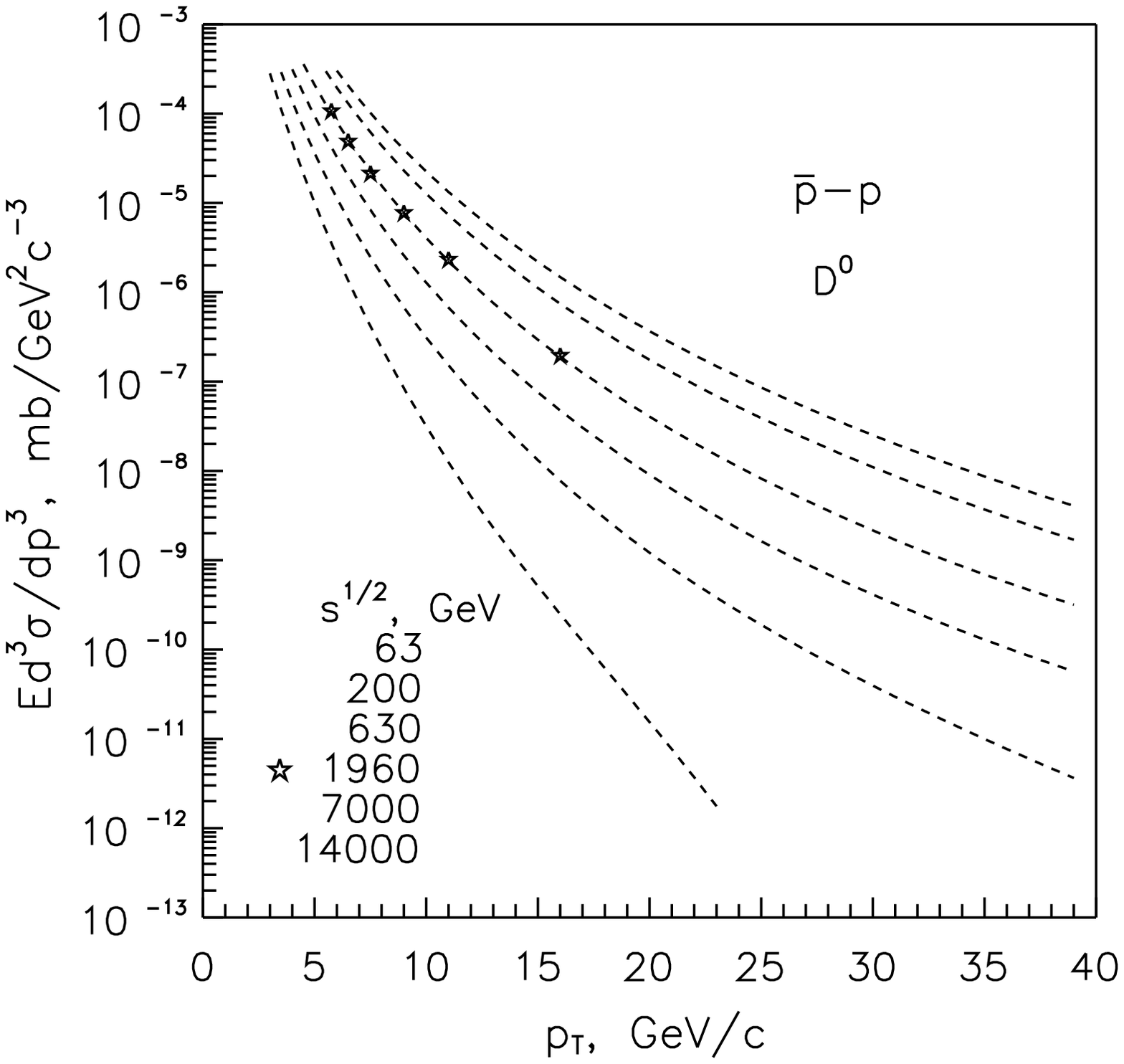}
\hspace*{40mm} (a) \hspace*{80mm} (b)
\caption{Spectra of $D^0$ mesons produced in $\bar{p}p$ collisions
in $z$ and $p_T$ presentations.
Experimental data are taken from \cite{Banner,D0_meson}.}
\end{figure}
The hypothesis of the flavor and energy  independence of $z$-scaling
for $D^0$ meson production was used in the analysis.
The data \cite{Banner} on inclusive cross section
of $\pi^0$-mesons produced in $\bar{p}p$ collisions at $\sqrt s =540$~GeV
and  the CDF data \cite{D0_meson} were used
for construction of the scaling function of $D^0$ mesons
over a wide range of $z$. The results of analysis are shown in Fig. 13(a).
The dashed lines are obtained by fitting the data \cite{Banner,D0_meson}
in $z$  presentation.
Figure 13(b) demonstrates transverse spectra of prompt $D^0$ meson production
in $\bar{p}p$ collisions over the kinematical
range $\sqrt s = 63-14000$~GeV, $p_T=5-40$~GeV/c
and $\theta_{cms}=90^0$ predicted by $z$-scaling.
The obtained results is of interest for comparison with
QCD predictions and experimental data over
a wider range of $p_T$.

{\subsection{$B^+$-mesons}}

Results of the first direct measurements of the $B$ meson
differential cross sections in $\bar{p}p$ collisions at $\sqrt s = 1800$~GeV
by measuring the mass and momentum of the $B$ meson decaying
into exclusive final states were presented in \cite{B+meson}.
The cross section was measured in the central rapidity region $|y|<1$
for $p_T(B)>6.0$~GeV/c.
Here we analyze in the framework of $z$-scaling the data
on transverse spectrum of $B^+$ mesons reconstructed
via the decay $B^+\rightarrow J/\psi K^+ $ with $J/\psi\rightarrow \mu^+\mu^-$.

\begin{figure}
\hspace*{10mm}
\includegraphics[width=60mm,height=60mm]{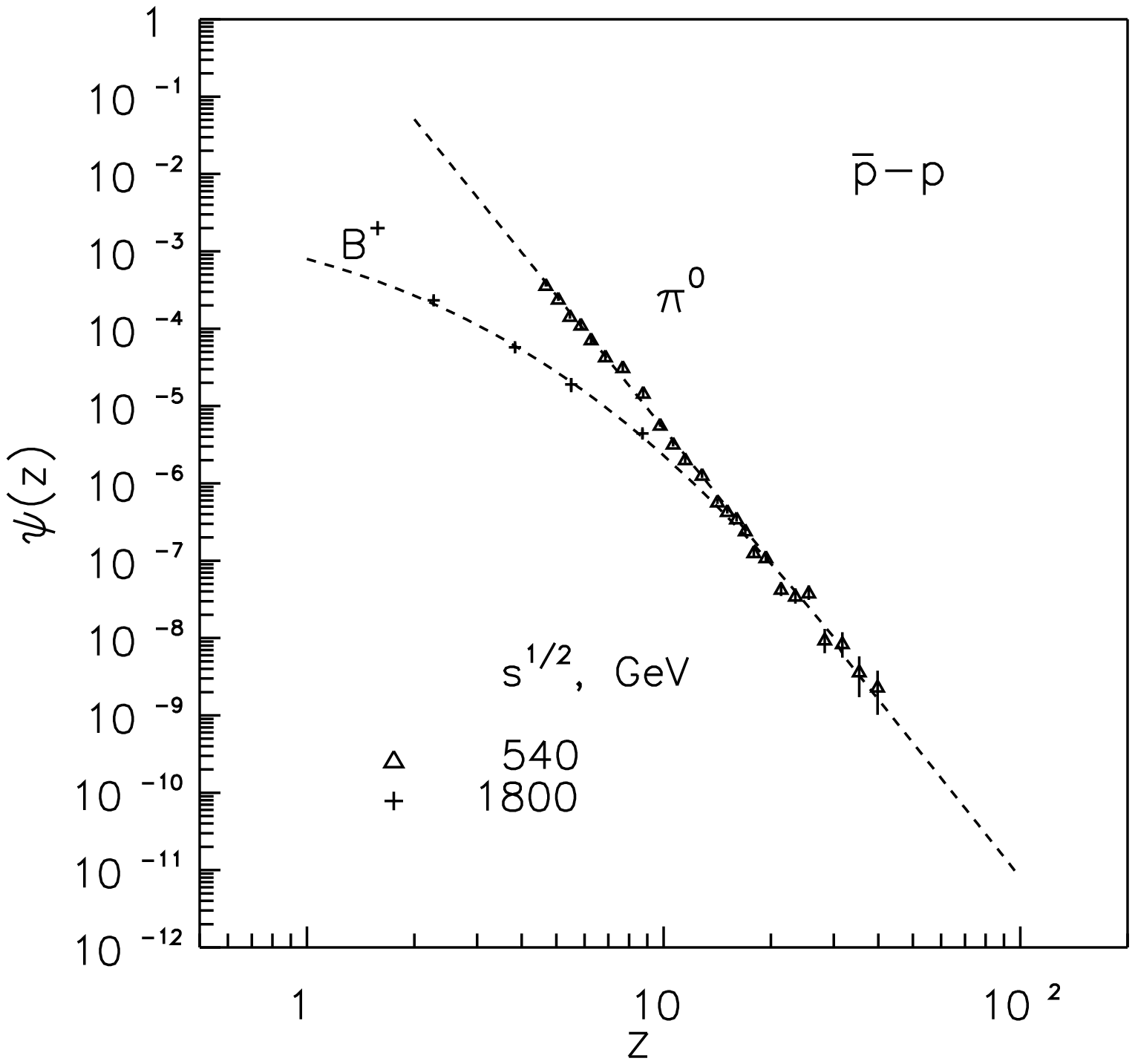}
\hspace*{15mm}
\includegraphics[width=60mm,height=60mm]{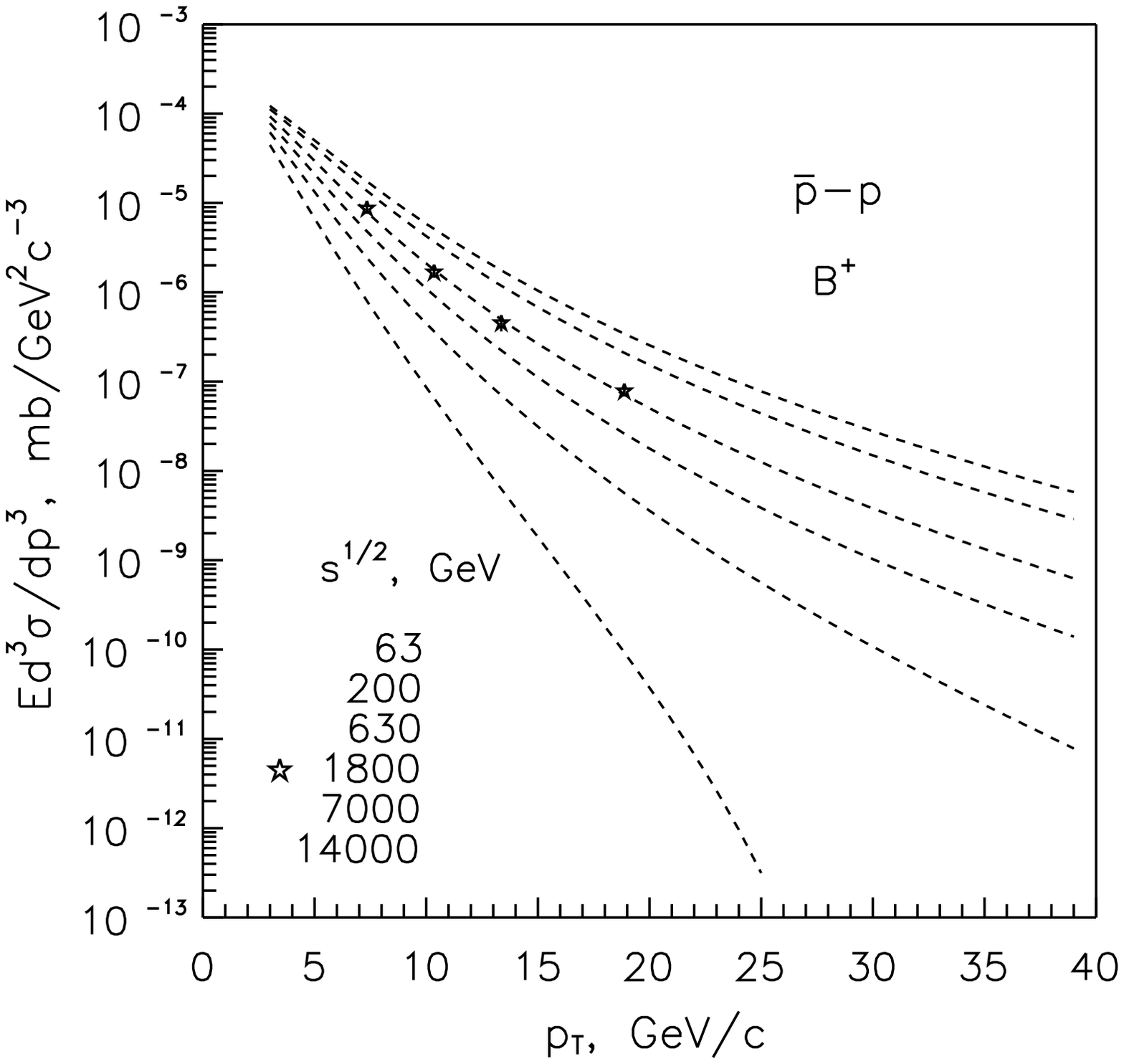}
\hspace*{40mm} (a) \hspace*{80mm} (b)
\caption{ Spectra of $B^+$ mesons produced
in $\bar{p}p$ collisions in  $z$ and $p_T$ presentations.
 Experimental data are taken from \cite{Banner,B+meson}.}

\end{figure}

Figure 14(a) shows the experimental data on inclusive spectra
for $B^+$ \cite{B+meson} and $\pi^+$ \cite{Banner} mesons produced
in $\bar{p}p$ in $z$  presentation which are used to construct the scaling function $\psi(z)$
for low and high $z$. The predictions of inclusive cross sections
of $B^+$ mesons over a range  $\sqrt s = 63-14000$~GeV, $\theta_{cms}=90^0$
and $p_T=5-40$~GeV/c are plotted in Fig. 14(b).
As seen from Fig. 14 the data  of transverse spectra
for $p_T>25$~GeV/c are necessary to determine
the overlapping range and study the asymptotic behavior of $\psi(z)$.
Note also that deviation of the scaling function
from the power law for low $z$ enhances with
mass of produced particle (see Figs. 14(a) and 13(a)).

{\subsection{$\Upsilon(1S)$-mesons}}

Features of heavy flavor production in high energy hadron collisions
at high $p_T$ can be related to new physics phenomena at small scales.
Understanding of flavor origin and search for similarity of
particle properties depending on the additive quantum numbers
(strangeness, charm, beauty, top)
is fundamental problem of particle physics.

The differential cross sections of $\Upsilon$ production in $\bar{p}p$ collisions
in the rapidity range $|y|<0.4$ in $1S, 2S$ and $3S$ states at $\sqrt s =1800$~GeV
are presented in \cite{Ups}. The three resonances were reconstructed through
the decay $\Upsilon\rightarrow \mu^+\mu^-$.
Transverse momentum of $\Upsilon$ was
measured over the range $p_T=0.5-20$~GeV/c.
The shape of $p_T$ spectrum  was found to be the same for all states.
The data were noted to be important for the investigation
of the bound state production mechanisms in $\bar{p}p$ collisions.

\begin{figure}
\hspace*{10mm}
\includegraphics[width=60mm,height=60mm]{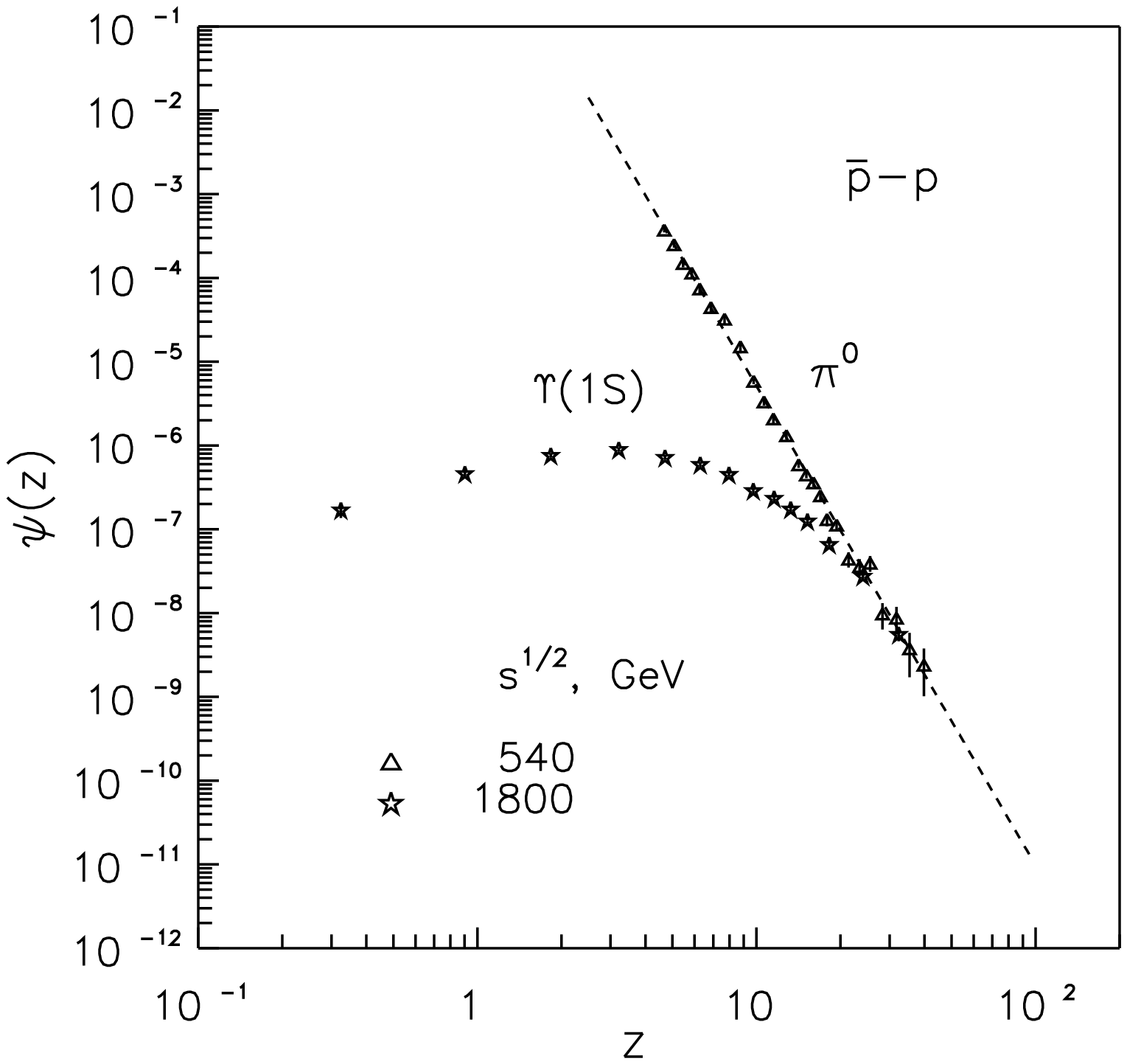}
\hspace*{15mm}
\includegraphics[width=60mm,height=60mm]{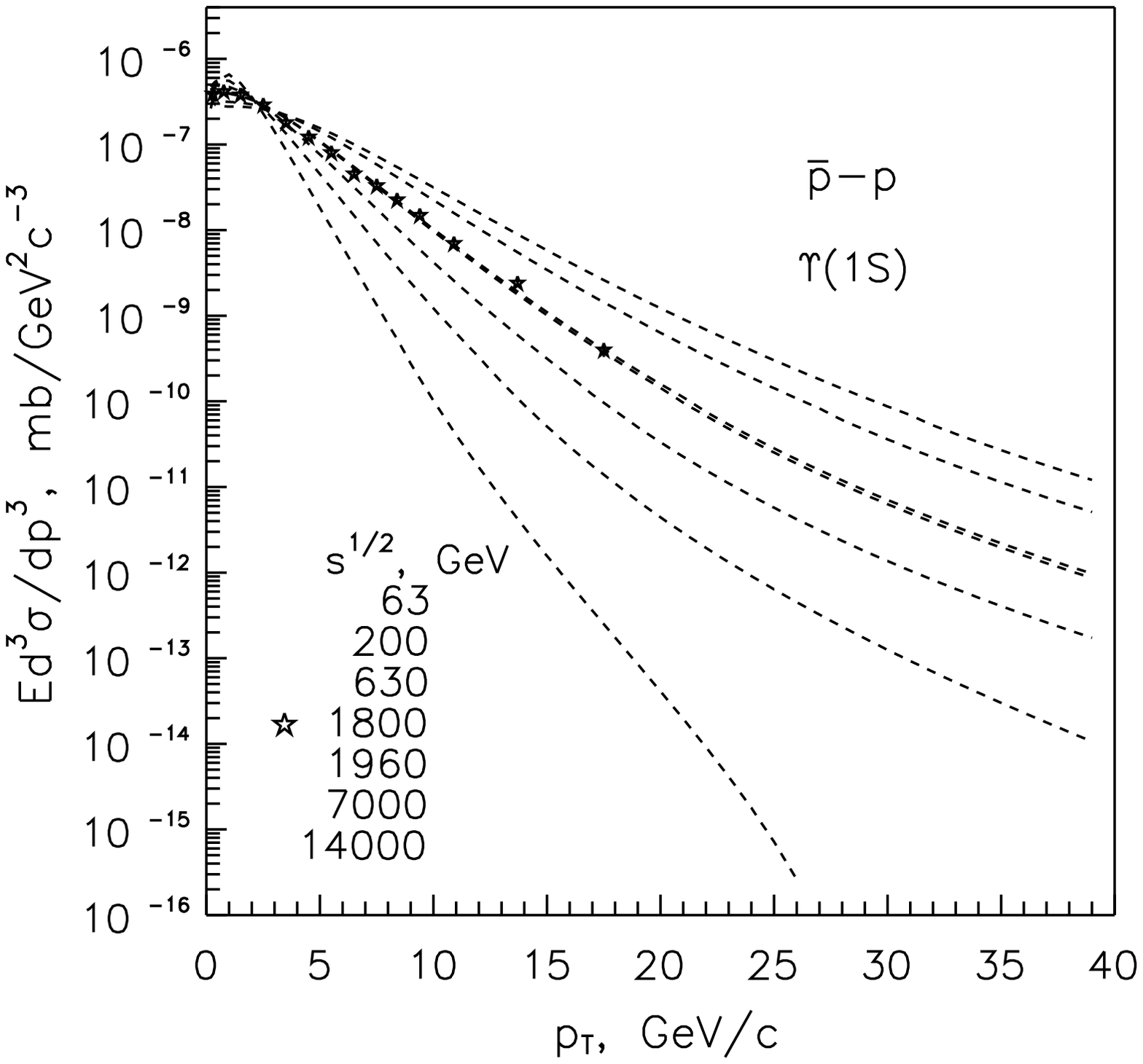}
\hspace*{45mm} (a) \hspace*{75mm} (b)
\caption{ Spectra of $\Upsilon(1S)$ mesons produced
 in $\bar{p}p$ collisions in  $z$ and $p_T$ presentations.
Experimental data are taken from \cite{Banner,Ups}.}
\end{figure}

Figure 15 demonstrates results of our analysis
of experimental data of $\Upsilon (1S)$ production
in $\bar{p}p$ collisions.
The scaling function of $\Upsilon (1S)$ (see Fig.15(a)) was constructed
using both the $\Upsilon (1S)$ \cite{Ups}
and $\pi^0$ \cite{Banner} transverse spectra.
For low $z$ the scaling function $\psi(z)$ deviates very strongly
from the power law shown by the dashed line.
The slope parameter of $\psi(z)$ changes the sign with positive on negative
at $z\simeq 3$.
Our predictions of inclusive cross sections over a range $\sqrt s = 63-14000$~GeV,
$\theta_{cms}=90^0$ and $p_T=1-40$~GeV/c are shown in Fig. 15(b).
Comparison of the obtained spectra with experimental data and QCD results
are of interest for verification of the flavor independence of $z$-scaling
and test of mechanisms of vector meson production described in the QCD theory.

{\subsection{$Z^0, W^+$ bosons}}

Vector $Z$ and $W$ bosons are carriers
of electroweak and strong interactions.
Therefore high precise data on transverse spectra
for both $Z$ and $W$ bosons produced in high energy
$\bar{p}p$ collisions can provide an important tests
and a direct confirmation of the unified model of
the weak, electromagnetic and strong interactions
(the Standard Model).
Quantum chromodynamics ascribes the transverse momentum
of the vector bosons produced in $\bar{p}p$ collisions
to associated production of one or more gluons or quarks with the boson.
Therefore data on the differential cross sections
for boson production can provides an important
test of our understanding of mechanism of boson
production described in the framework of QCD.
The large mass of the vector bosons assures a large energy scale
for probing perturbative QCD with good reliability.
It also provides bounds on parametrizations of the parton distribution functions
used to describe the nonperturbative regime of QCD processes.
Deviations from the prediction for  high $p_T$ could indicate
new physics phenomena beyond the Standard Model.

Measurements of the differential cross sections for $Z$ and $W$
boson production as a function of transverse momentum
in $\bar{p}p$ collisions at $\sqrt s =1800$~GeV over a range
$p_T=1-200$~GeV/c are presented in \cite{Z0,W+}.
The $W$ and $Z$ bosons were detected through their
leptonic decay modes ($W\rightarrow e\nu, Z\rightarrow e^+e^-$).

We assume that asymptotic behavior of the scaling function
of vector bosons and direct photons produced in $\bar{p}p$
collisions is similar.
It describes by the power law for high $z$
with the same value of the slope parameter $\beta $ of $\psi(z)$
for both particles.
Direct confirmation of the similarity  of direct and virtual photon (Drell-Yan pair)
and vector boson production for high $p_T$ is considered as an important
feature of constituent interactions.

The asymptotic behavior of $\psi(z)$ of direct photons
was used to construct $\psi(z)$ of vector bosons over  a wide range of $z$.
Figures 16(a) and 17(a) demonstrate $z$  presentation of data
on inclusive transverse spectra for vector bosons \cite{Z0,W+}
and direct photons \cite{Zg}.
Note that the overlapping range of $\psi(z)$ for the $W^+$ boson
and $\gamma$ is large enough. Nevertheless
the direct measurements of transverse spectra  of the $W^+$ bosons
for $p_T> 200$~GeV/c are necessary for verification of the assumption.
Figures 16(b) and 17(b) show the dependence of inclusive
cross sections of $Z$ and $W^+$ bosons produced in $\bar{p}p$ collisions
over a range $\sqrt s = 200-14000$~GeV, $p_T = 1-500$~GeV/c
and $\theta_{cms}=90^0$.



\begin{figure}
\hspace*{10mm}
\includegraphics[width=60mm,height=60mm]{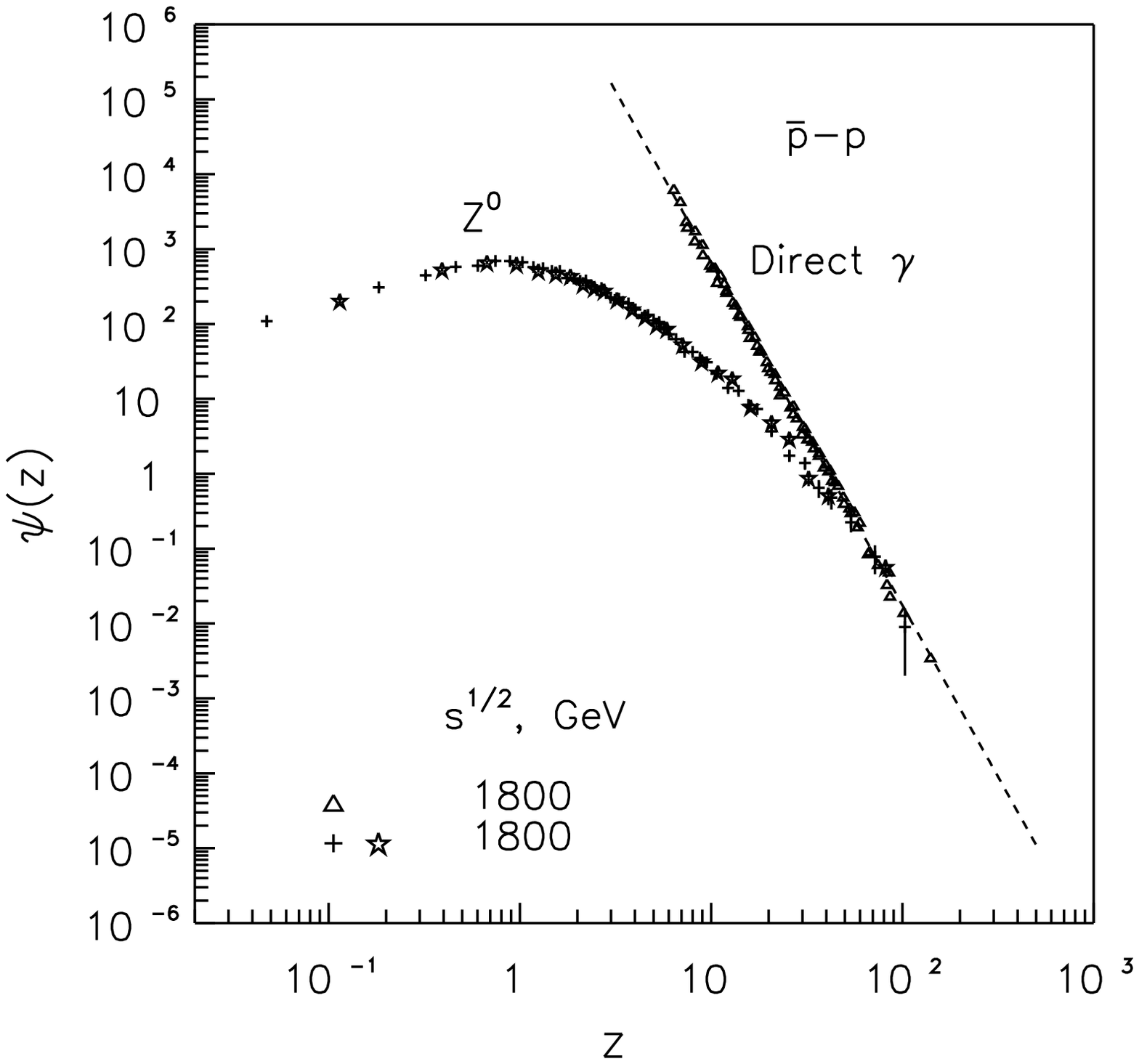}
\hspace*{15mm}
\includegraphics[width=60mm,height=60mm]{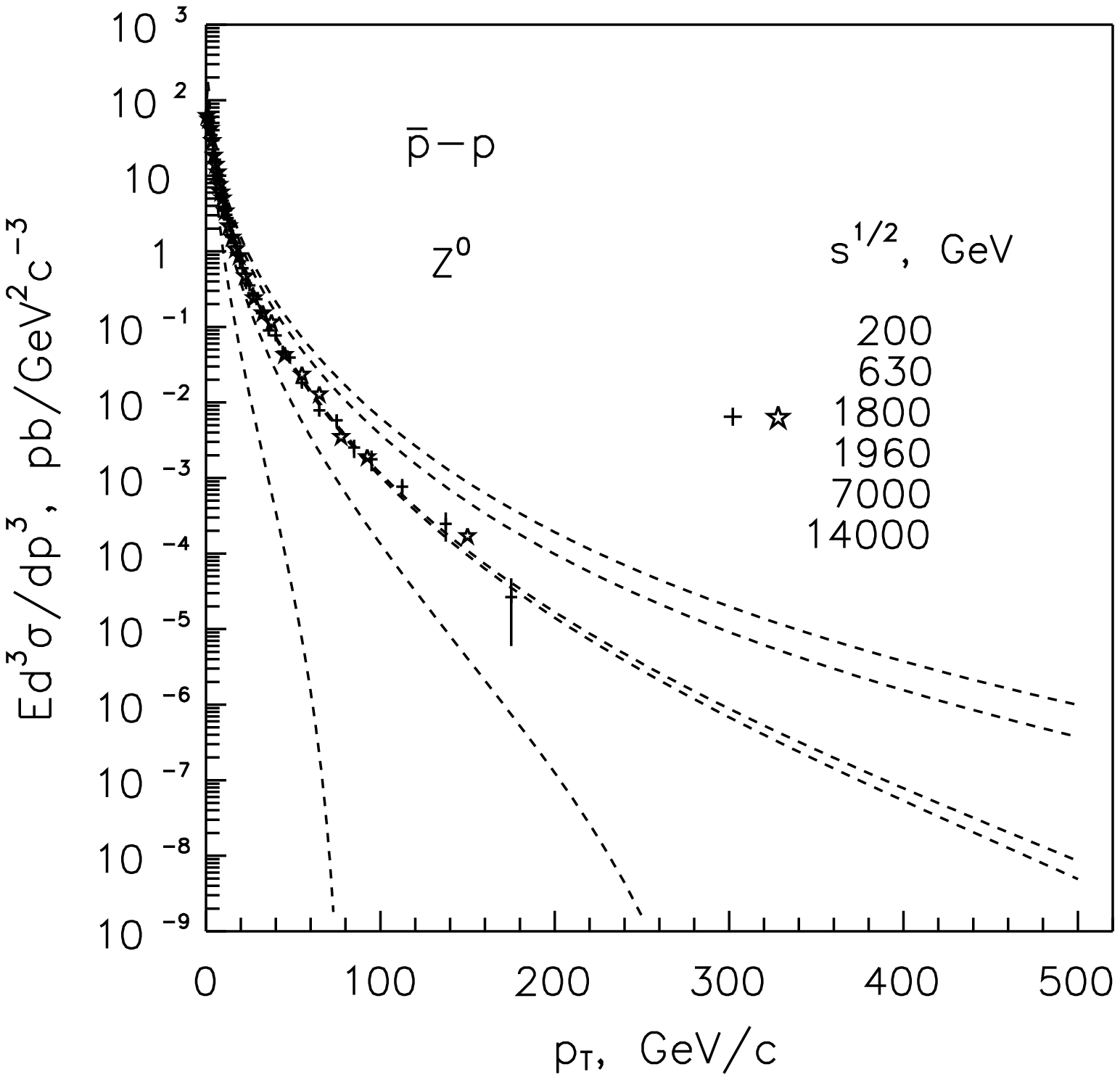}
\hspace*{45mm} (a) \hspace*{75mm} (b)
\caption{ Spectra of $Z^0$ bosons produced in $\bar {p}p$ collisions
in $z$ and $p_T$ presentations.
Experimental data are taken from \cite{Zg1,Z0}.}

\end{figure}


\begin{figure}
\hspace*{10mm}
\includegraphics[width=60mm,height=60mm]{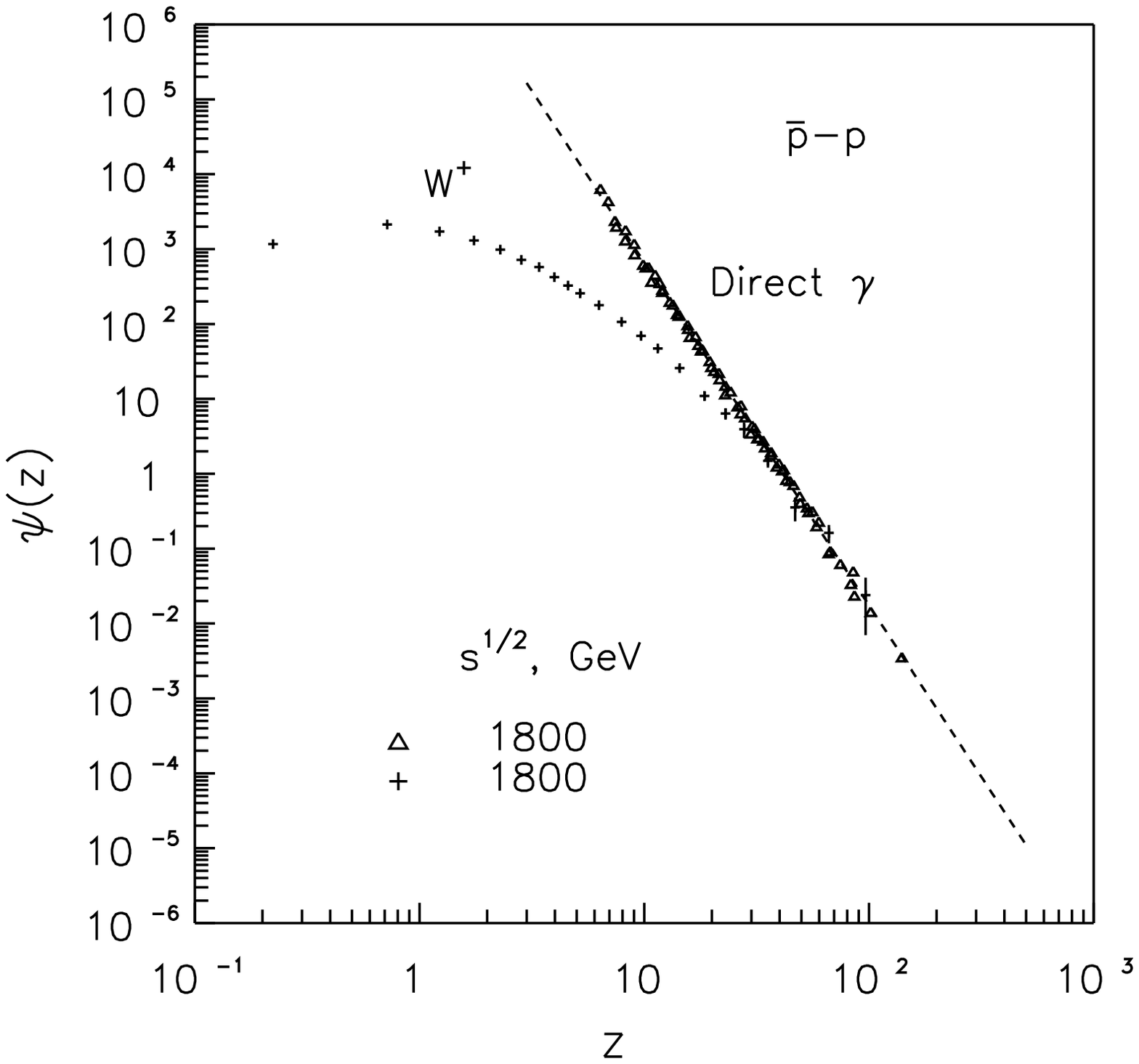}
\hspace*{15mm}
\includegraphics[width=60mm,height=60mm]{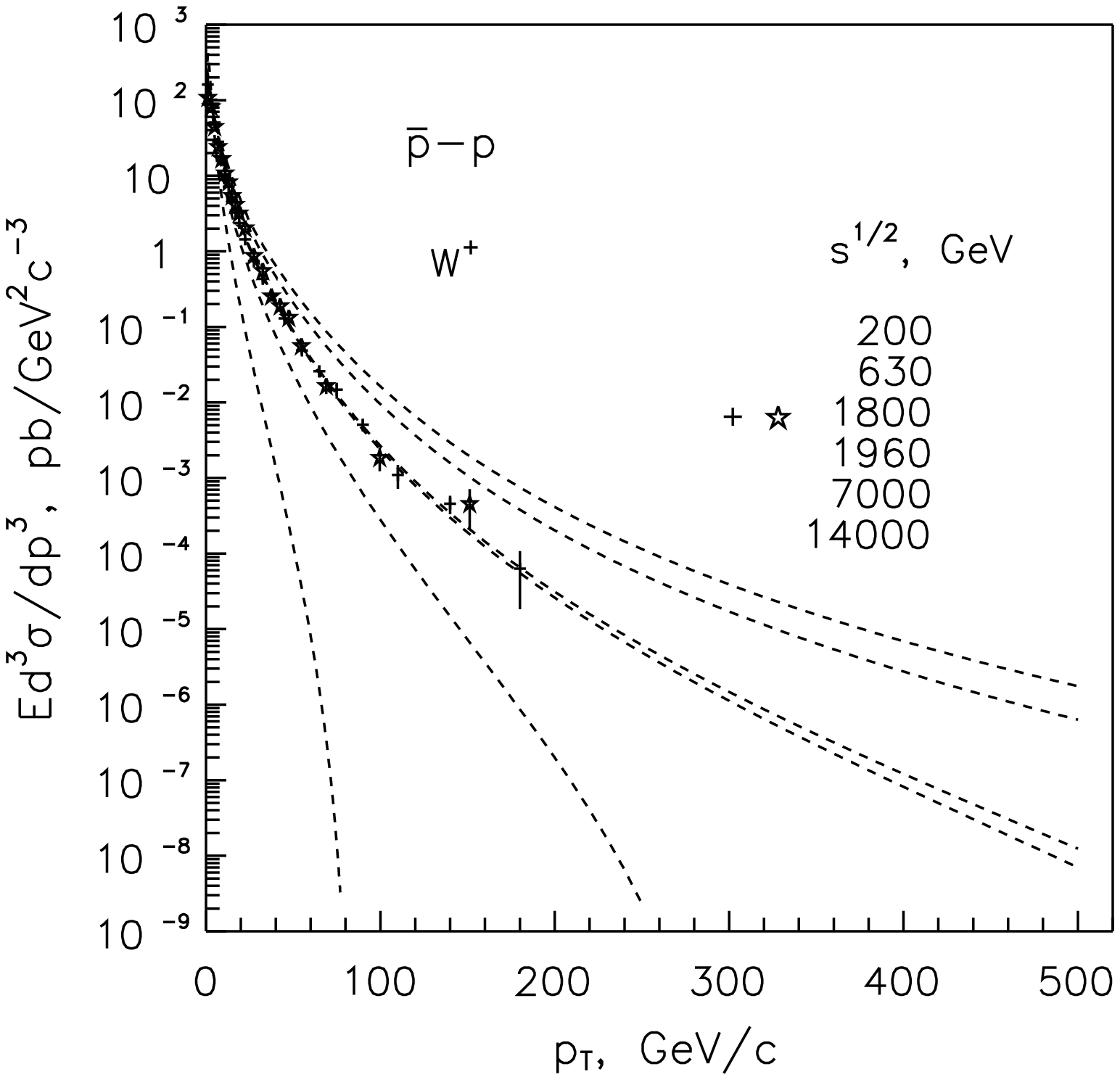}
\hspace*{45mm} (a) \hspace*{75mm} (b)
\caption{ Spectra of $W^+$ bosons produced in $\bar{p}p$ collisions
in $z$ and $p_T$ presentations.
Experimental data are taken from \cite{Zg1,W+}.}
\end{figure}

\section{Conclusions}

The results of recent analysis in the framework of $z$-scaling
of experimental data on inclusive cross sections of particle
production  in $pp, AA$ and $\bar{p}p$ collisions
at high $p_T$ at RHIC and Tevatron
were reviewed.
Physical concept of $z$-scaling and
interpretation of the function $\psi(z)$ and variable $z$
were  discussed. It was shown that the generalized concept
of $z$-scaling allows us to study the multiplicity dependence
of particle spectra and restore the multiplicity independence
of the scaling function.
The properties of $z$  presentation of experimental data were  verified.
We consider that the properties established in the present analysis of data
reflect general features of the structure
of the colliding objects, interaction of their constituents
and mechanism of particle formation.

The hypothesis on universality of asymptotic behavior
of the scaling function (for $\pi^0$-mesons, direct photons)
was used for construction of $\psi(z)$
for $J/\psi, \Upsilon(1S), D^0, B^+, Z, W^+$ particles
and predictions of transverse spectra
over a wide range of $\sqrt s $ and $p_T$.
The obtained results support main idea of $z$-scaling based
on property of self-similarity of particle interactions
for high $p_T$.

We consider that the $z$-scaling may be used as a tool
for searching for new physics phenomena beyond the Standard Model
in hadron and nucleus collisions at high transverse
momentum and high multiplicity at RHIC, Tevatron and LHC.

\vskip 5mm

{\bf Acknowledgments}.
The investigations have been partially supported by the IRP
AVOZ10480505, by the Grant Agency of the Czech Republic under
the contract No. 202/04/0793 and by the special program of
the Ministry of Science and Education of the Russian Federation,
grant RNP.2.1.1.5409.


\begin{thebibliography}{99}

\bibitem{Quarkcom}
E. Eichten, K. Lane, and M. Peskin, Phys. Rev. Lett. {\bf 50}, 811 (1983).\\
E. Eichten, I. Hinchliffe, K. Lane, and C. Quigg, Rev. Mod. Phys.
{\bf 56}, 4 (1984).

\bibitem{Extradim}
I. Antoniadis, in {\it Proceedings of European School of
High-Energy Physics}, Beatenberg, Switzerland, 26 August - 8
September, 2001 (Editors: N. Ellis and J. March-Russul) p.301.

\bibitem{Blackhole}
C.G. Lester, in {\it Proceedings of Advanced Studies Institute on
"Physics at LHC"}, Czech Republic, Prague, July 6-12, 2003
(Editors: M. Finger, A. Janata, and M. Virius) A303.

\bibitem{Fracspace}
L. Nottale, {\it Fractal Space-Time and Microphysics} (World Sci., Singapore, 1993).\\
B. Mandelbrot, {\it The Fractal Geometry of Nature} (Freeman, San Francisco, 1982).


\bibitem{RHIC_review}
I.~Arsene {\it et al.} (BRAHMS collaboration),
Nucl. Phys. {\bf A757}, 1 (2005).\\
B.B.~Back {\it et al.} (PHOBOS collaboration),
Nucl.Phys. {\bf A757}, 28 (2005).\\
J.~Adams {\it et al.} (STAR collaboration),
Nucl. Phys. {\bf A757}, 102 (2005).\\
K. Adcox {\it et al.} (PHENIX collaboration),
Nucl. Phys. {\bf A757}, 184 (2005).

\bibitem{Feynman}
R.P. Feynman, Phys. Rev. Lett. {\bf 23}, 1415 (1969).

\bibitem{Bjorken}
J.D. Bjorken, Phys. Rev. {\bf 179}, 1547 (1969).\\
 J.D. Bjorken, and E.A. Paschos, Phys. Rev. {\bf 185}, 1975 (1969).

\bibitem{Bosted}
P. Bosted {\it et al.}, Phys. Rev. Lett. {\bf 49}, 1380 (1972).

\bibitem{Benecke}
J. Benecke  {\it et al.},  Phys. Rev.  {\bf 188}, 2159 (1969).

\bibitem{Baldin}
A.M. Baldin,
Sov. J. Part. Nucl. {\bf 8}, 429 (1977).

\bibitem{Stavinsky}
 V.S. Stavinsky,
Sov. J. Part. Nucl. {\bf 10}, 949 (1979).

\bibitem{Leksin}
G.A. Leksin: Report No. ITEF-147, 1976; G.A. Leksin: in
{\it Proceedings of the XVIII International Conference on High
Energy Physics}, Tbilisi, Georgia, 1976, edited by N.N.
Bogolubov {\it et al.}, (JINR Report No. D1,2-10400, Tbilisi,
1977), p. A6-3.

\bibitem{KNO}
Z. Koba, H.B. Nielsen, and P. Olesen,
Nucl. Phys.  {\bf B40}, 317 (1972).

\bibitem{Matveev}
  V.A. Matveev, R.M. Muradyan, and A.N. Tavkhelidze,
  Part. Nuclei {\bf 2}, 7 (1971);
  Lett. Nuovo Cim. {\bf 5},  907 (1972);
  Lett. Nuovo Cim. {\bf 7}, 719 (1973).

\bibitem{Brodsky}
  S. Brodsky, and G. Farrar,
  Phys. Rev. Lett. {\bf 31}, 1153 (1973);
  Phys. Rev. D {\bf 11}, 1309 (1975).



\bibitem{Z}
I. Zborovsk\'{y}, Yu.A. Panebratsev, M.V. Tokarev, and G.P. \v{S}koro,
Phys. Rev. D {\bf 54}, 5548 (1996); I. Zborovsk\'{y}, M.V. Tokarev,
Yu.A. Panebratsev, and G.P. \v{S}koro, Phys. Rev. C {\bf 59}, 2227
(1999);
 M.V. Tokarev,
O.V. Rogachevski, and T.G. Dedovich, Preprint No. E2-2000-90, JINR
(Dubna, 2000); M. Tokarev, I. Zborovsk\'{y}, Yu. Panebratsev, and
G. Skoro, Int. J. Mod. Phys. {\bf A16}, 1281 (2001); M. Tokarev,
hepph/0111202; M. Tokarev and D. Toivonen, hepph/0209069;
G.P. Skoro, M.V. Tokarev, Yu.A. Panebratsev, and I. Zborovsk\'{y},
hepph/0209071; M. Tokarev, Acta Physica Slovaca, {\bf 54}, 321 (2004).

\bibitem{Zpi0} M.V. Tokarev, O.V. Rogachevski, and T.G. Dedovich,
 J. Phys. G: Nucl. Part. Phys. {\bf 26}, 1671 (2000).

\bibitem{Zg}
M. Tokarev, G. Efimov, hep-ph/0209013.\\
M.V. Tokarev, G.L. Efimov, and D.E. Toivonen,
Physics of Atomic Nuclei, {\bf 67}, 564 (2004).

\bibitem{Zjet}
M.V. Tokarev, and T.G. Dedovich, Int. J. Mod. Phys. {\bf A15}, 3495 (2000).\\
M.V. Tokarev, and T.G. Dedovich, Physics of Atomic Nuclei, {\bf 68}, 404 (2005).


\bibitem{ZZ}
I. Zborovsk\'{y}, and M.V. Tokarev,  Part. and Nucl., Letters,
{\bf 3}, 68 (2006); hep-ph/0506003.


\bibitem{Z3}
I.~Zborovsk\'{y}, and  M.V.~Tokarev, these proceedings.

\bibitem{ZS}
M.V.~Tokarev, and I.~Zborovsk\'{y},
In: Proceedings of XXXVI International Symposium
on Multiparticle Dynamics (ISMD2006),
September 02-08, 2006, Paraty, Rio de Janeiro, Brazil;
http://www.sbf1.sbfisica.org.br/eventos/extras/ismd2006.


\bibitem{ZZZ}
I.~Zborovsk\'{y}, and  M.V.~Tokarev, JINR Commun. E2-2006-34, Dubna, 2006, 20p.

\bibitem{pi0_PHENIX}
H.~Buesching, (PHENIX Collaboration)
"Hot Quark$'$06",  May 15-20, 2006,  Sardinia, Italy.
\bibitem{pi0_STAR}
D.~Relyea, (STAR Collaboration)  RHIC \& AGS Annual Users Meeting,
   June 8-9, 2006,  BNL, USA.

\bibitem{Angel} A.L.S. Angelis {\it et al.},
  Phys. Lett. {\bf B79}, 505 (1978).

\bibitem{Kou1} C. Kourkoumelis {\it et al.},
 Phys. Lett. {\bf B83}, 257 (1979).

\bibitem{Kou3} C. Kourkoumelis {\it et al.},
  Z. Phys. {\bf 5}, 95 (1980).

\bibitem{Lloyd} D. Lloyd Owen  {\it et al.},
 Phys. Rev. Lett. {\bf 45}, 89 (1980).

\bibitem{Eggert} K. Eggert {\it et al.},
  Nucl. Phys. {\bf B98}, 49 (1975).


\bibitem{SFM} A.~Breakstone {\it et al.} Z. Phys. {\bf C33}, 333 (1987).

\bibitem{mult_UA1}
   G.~Arnison  {\it et al.}, (UA1 collaboration), Phys. Lett.  {\bf B118}, 167 (1982).
\bibitem{E735_pT}
  T.~Alexopoulos  {\it et al.} (E735 collaboration),
  Phys. Rev. Lett. {\bf 60}, 1622  (1988).

\bibitem{mult_E735}
T.~Alexopoulos  {\it et al.} (E735 collaboratoion), Phys. Lett. {\bf B336}, 599 (1994).

\bibitem{mult_STAR}
J.E.~Gans,  PhD Thesis, Yale University, USA (2004).

\bibitem{mult_CDF}
D.~Acosta  {\it et al.}, (CDF collaboration) Phys. Rev. {\bf D65}, 072005 (2002).\\
F.~Rimondi {\it et al.,} Yad. Fiz.  {\bf 67}, 128 (2004).

\bibitem{Witt}
R.~Witt (STAR collaboration), J. Phys. G: Nucl. Part. Phys. {\bf 31}, S863 (2005).\\
B.I.~Abelev {\it et al.,} nucl-ex/0607033.

\bibitem{STAR_jet}
M.~Miller (STAR collaboration), hep-ex/0604001.\\
B.I.~Abelev {\it et al.}, Phys. Rev. Lett. 97,  252001 (2006); hep-ex/0608030.

\bibitem{Ded_Tok} T.G.~Dedovich and M.V.Tokarev, these proceedings.

\bibitem{STAR200}
J.~Adams {\it et al.} (STAR collaboration), Phys. Rev. Lett. {\bf 91}, 172302 (2003).

\bibitem{PHOBOS_Cu}
B.B.~Back {\it et al.} (PHOBOS collaboration),  Phys. Lett. {\bf  B578}, 297 (2004).\\
B.~Alver {\it et al.} (PHOBOS collaboration),
Phys. Rev. Lett. {\bf 96},  212301 (2006).\\
E.~Wenger (PHOBOS collaboration), nucl-ex/0511036.

\bibitem{D0_g}
A.~Abachi {\it et al.} (D0 collaboration), Phys.Rev.Lett. {\bf 77}, 5011 (1996).\\
B.~Abbott {\it et al.} (D0 collaboration), Phys.Rev.Lett. {\bf 84}, 2786 (2000).\\
V.M.~Abazov {\it et al.} (D0 collaboration), Phys. Lett. {\bf B639}, 151 (2006).

\bibitem{Zg1}
 A.~Abachi {\it et al.} (D0 collaboration), Phys. Rev. Lett. {\bf 77}, 5011 (1996).\\
 F.~Abe {\it et al.} (CDF collaboration), Phys. Rev. Lett. {\bf 68}, 2734 (1992).\\
 F.~Abe {\it et al.} (CDF collaboration), Phys. Rev. {\bf D48}, 2998 (1993).


\bibitem{D0_CDF_jet}
M.~D${'}$Onofrio (for CDF \& D0 collaborations),
XX Rencontres de Physique de la Vallee D${'}$Aoste,
La Tuile, Italy, March 5-11, 2006;
http://www.pi.infn.it/lathuile/2006/Programme.htm.\\
D.~Bandurin,   (for CDF \& D0 collaborations),
9th Conference on the Intersections of Nuclear
and Particle Physics (CIPANP$'$06), Rio Grande, Puerto Rico,
May 30 – June 4, 2006; http://cipanp.physics.uiuc.edu/.

\bibitem{Matthias}
M.~T$\rm {\ddot o}$nnesmann, Eur. Phys. J. {\bf C33}, s422 (2004).

\bibitem{Zflavor} M.~Tokarev,
In: Proceedings of the International Workshop
"Relativistic Nuclear Physics: from Hundreds of MeV to Tev",
Varna, Bulgaria, September 10-16, 2001
(Dubna, JINR, E1,2-2001-290, 2001, 300 p.), V.1, p.280-300.

\bibitem{bjet}
B.~Abbott {\it et al.} (D0 collaboration), Phys. Rev. Lett. {\bf 85}, 5068 (2000).\\
M.~D${'}$Onofrio (for D0 \& CDF collaborations),
 XXXXth Rencontres de Moriond,  "QCD and Hadronic interactions at high energy",
2-19 March, 2005, Italy; http://moriond.in2p3.fr/QCD/2005/

\bibitem{Tok_Jpsi} M.~Tokarev,  hep-ph/0405230.

\bibitem{Jpsi1}
F. Abe {\it et al.} (CDF collaboration),   Phys. Rev. Lett. {\bf 69}, 3704 (1992).\\
F. Abe {\it et al.} (CDF collaboration),   Phys. Rev. Lett.  {\bf 79}, 572 (1997).\\
F. Abe {\it et al.} (CDF collaboration),   Phys. Rev. Lett.{\bf 79}, 578 (1997).

\bibitem{Jpsi2}
Yu.~Gotra, Ph.D. Thesis (JINR, Dubna, Russia, 2004).\\
D.~Acosta {\it et al.} (CDF collaboration), Phys. Rev. {\bf D71},  032001 (2005).\\
D.~Acosta {\it et al.} (CDF collaboration),  Phys. Rev. Lett. {\bf 96}, 202001 (2006).

\bibitem{Banner}
M.~Banner {\it et al.},  Phys. Lett. {\bf B115}, 59 (1982).

\bibitem{D0_meson}
D.~Acosta {\it et al.} (CDF collaboration), Phys. Rev. Lett. {\bf 91}, 241804 (2003).


\bibitem{B+meson}
F.~Abe {\it et al.} (CDF collaboration),  Phys. Rev. Lett. {\bf 75}, 1451 (1995).\\
D.~Acosta {\it et al.} (CDF collaboration), Phys. Rev. {\bf D65},  052005 (2002).



\bibitem{Ups}
D.~Acosta {\it et al.} (CDF collaboration), Phys. Rev. Lett. {\bf 88}, 161802  (2002).


\bibitem{Z0}
 T.~Affolder {\it et al.} (CDF collaboration), Phys. Rev. Lett. {\bf 84}, 845 (2000).\\
 F.~Abe {\it et al.} (CDF collaboration),  Phys. Rev. Lett. {\bf 67}, 2937 (1991).\\
 B.~Abbott {\it et al.} (D0 collaboration), Phys.Rev. {\bf D61}, 032004 (1999).\\
 V.M.~Abazov {\it et al.} (D0 collaboration), Phys. Lett. {\bf B517}, 299 (2001).

\bibitem{W+}
F.~Abe  {\it et al.} (CDF collaboration),  Phys. Rev. Lett. {\bf 66}, 2951 (1991).\\
V.M.~Abazov  {\it et al.} (D0 collaboration), Phys. Lett. {\bf B517}, 299 (2001).\\
B.~Abbott {\it et al.} (D0 collaboration),  Phys. Lett. {\bf B513}, 292 (2000).



\end{thebibliography}
\end{document}